\documentclass[aps,showpacs]{revtex4}

\usepackage{graphicx}
\usepackage{dcolumn}
\usepackage{bm}

\begin{document}


\title{Unified model for network dynamics exhibiting nonextensive statistics}

\author{Stefan Thurner $^{1,}$}
\email{thurner@univie.ac.at}
\author{Fragiskos Kyriakopoulos$^{1,2}$}
\author{Constantino Tsallis$^{3,4,}$}
\email{tsallis@santafe.edu} 
\affiliation{
   $^1$Complex Systems Research Group, 
       HNO Medical University of Vienna, W\"ahringer G\"urtel 18-20, A-1090, Austria \\
   $^2$Institute of Theoretical Physics, 
      Johannes Kepler University, Altenbergerstrasse 69, Linz, A-4040, Austria \\
   $^{3}$Santa Fe Institute, 1399 Hyde Park Road, Santa Fe, NM 87501, USA \\
   $^{4}$Centro Brasileiro de Pesquisas Fisicas,
        Rua Xavier Sigaud 150, 22290-180 Rio de Janeiro-RJ; Brazil}

\date{\today}

\begin{abstract}
We introduce a dynamical network model which unifies a number of network families 
which are individually known to exhibit $q$-exponential degree distributions. The 
present model dynamics incorporates static (non-growing) self-organizing networks, 
preferentially  growing networks, and  (preferentially) rewiring networks. 
Further, it exhibits a natural random
graph limit. The proposed model generalizes network dynamics to rewiring and 
growth modes which depend on internal topology as well as on a metric imposed by
the space they are embedded in.  In all of the networks emerging from the presented
model we find $q$-exponential degree distributions over a large parameter space.  
We comment on the parameter dependence of the corresponding entropic index $q$ 
for the degree distributions, and on the behavior of the clustering coefficients  
and neighboring connectivity distributions.
\end{abstract}



\maketitle

\section{Introduction}
Over the past two decades, {\it nonextensive
statistical mechanics} has successfully addressed a wide spectrum of nonequilibrium phenomena in non-ergodic and other complex systems \cite{tsallis88,gellmann}. Recently, it has also  entered the field of networks
\cite{tsallis_sm,thutsall,soares2,thurner05,hasegawa,wilk,white,latora}. 
Nonextensive statistical mechanics is a generalization of Boltzmann-Gibbs (BG)
statistical mechanics. It is based on the entropy 
\begin{equation}
S_q \equiv \frac{1-\int dx \, [p(x)]^q}{q-1} \;\;\;\;\;
\Bigl(
S_1=S_{BG} \equiv -\int dx \, p(x) \ln p(x)  \Bigr) \,.
\end{equation}
The extremization of the entropy $S_q$ under
appropriate constraints \cite{TsallisMendesPlastino} yields the stationary-state distribution. This is of the
$q$-exponential form, where the $q$-exponential function 
is defined as
\begin{equation}
e_{q}^x \equiv [1+(1-q)x]^{1/(1-q)} \quad , \label{qexp} 
\end{equation}
for $1+(1-q)x \ge 0$, and zero otherwise (with $e_1^x=e^x$). 
The tail exponent $\gamma \equiv 1/(q-1)$ characterizes the asymptotic power-law distribution.

Since the very beginning of the tremendous recent modeling efforts
of complex networks it has been noticed that degree distributions
asymptotically follow power-laws \cite{albert99}, or even exactly
$q$-exponentials \cite{albert00}. The model in \cite{albert99}
describes growing networks with a so-called preferential attachment
rule, meaning that any new node $i$ being added to the system links
itself to an already existing node $j$ in the network with a
probability that is proportional to the degree $k_j$ of node $j$. In
\cite{albert00} this model was extended to also allow for
preferential rewiring. The analytical solution to the model has a
$q$-exponential as a result, with the nonextensivity parameter $q$
being fixed uniquely by the model parameters. Recently in
\cite{tsallis_sm} preferential attachment networks have been
embedded in Euclidean space, where the attachment probability for a
newly added node is not only proportional to the degrees of existing
nodes, but also depends on the Euclidean distance between nodes. The
model is realized by setting the linking probability of a new node
to an existing  node $i$ to be $p_{\rm link}\propto
k_i/r_i^{\alpha}$ $(\alpha\geq0)$, where $r_i$ is the distance
between the new node and node $i$; $\alpha=0$ corresponds to the
model in \cite{albert99} which has no metrics. The analysis of 
the degree distributions of the resulting networks has exhibited \cite{tsallis_sm} 
$q$-exponentials with a clear $\alpha$-dependence of the
nonextensivity parameter $q$. In the large $\alpha$ limit, $q$
approaches unity, i.e., random networks are recovered in the
Boltzmann-Gibbs limit. In an effort to understand the evolution of
socio-economic networks, a model was proposed in \cite{white} that
builds upon \cite{albert00} but introduces  a rewiring scheme which
depends on the {\it internal} network distance between two nodes,
i.e., the number of steps needed to connect the two nodes. The
emerging degree distributions have been subjected to a statistical
analysis where the (null) hypothesis of $q$-exponentials could not be
rejected.

It has been found that networks exhibiting degree
distributions compatible with $q$-exponentials are not at all
limited to growing and preferentially organizing networks. A model
for nongrowing networks which was recently put forward in
\cite{thutsall} also unambiguously exhibits $q$-exponential degree
distributions. This model was motivated by interpreting networks as
a certain type of 'gas' where upon an (inelastic) collision of two
nodes,  links get transfered in analogy to the energy-momentum
transfer in real gases. In this model a fixed number of nodes in an
(undirected) network can 'merge', i.e., two nodes fuse into one
single node, which keeps the union of links of the two original
nodes; the link connecting the two nodes before the merger is
removed. At the same time a new node is introduced to the system and
is linked randomly to any of the existing nodes in the network
\cite{sneppen}. Due to the nature of this model the number of links
is not strictly conserved -- which can be thought of as jumps
between discrete states in some 'phase space'. The model has been
further generalized to exhibit a distance dependence as in
\cite{tsallis_sm}, however $r_i$ not being Euclidean but internal
distance. Again, the resulting degree distributions have
$q$-exponential form.

A quite different approach was taken in \cite{abe05} where an
ensemble interpretation of random networks has been adopted,
motivated by superstatistics \cite{beck}. Here it was assumed that
the average connectivity $\bar k$ 
in random networks is fluctuating
according to a distribution $\Pi(\bar k)$, which is sometimes
associated with a 'hidden-variable' distribution. In this sense a
network with any degree distribution can be seen as a
'superposition' of random networks with the degree distribution
given by $p(k)= \int_0^{\infty} d\lambda \, \Pi(\bar k)
\frac{\bar k ^k e^{-\bar k}}{k!}$. It was shown in \cite{abe05}, as
an exact example, that an asymptotically power-law functional form
of $\Pi(\bar k)\propto \bar k ^{-\gamma}$ leads to degree
distributions of Zipf-Mandelbrot form, $p(k) \propto
\frac{1}{(k_0+k)^{\gamma}}$, which is equivalent to a
$q$-exponential $e_q^{-k/\kappa}$ with $\kappa \equiv (1-q)k_0$ and $q \equiv 1+1/\gamma$. Very recently
a possible connection between {\it small-world} networks and the
maximum $S_q$-entropy principle, as well as to the hidden variable
method \cite{abe05}, has been noticed in \cite{hasegawa}.

In yet another view, networks have recently been treated as
statistical systems on a Hamiltonian basis
\cite{manna,vicsek,newman_sm,biely}. It has been shown that these
systems show a phase transition like behavior \cite{vicsek}, along
which networks structure changes. In the low temperature phase one
finds networks of 'star' type, meaning that a few nodes are
extremely well connected resulting even in a discontinuous $p(k)$;
in the high temperature phase one finds random networks.
Surprisingly, for a special type of Hamiltonians networks with
$q$-exponential degree distributions emerge right in the vicinity of
the transition point \cite{biely}.

Given the above characteristics of networks and the fact that
a vast number of real-world and model networks show asymptotic
power-law degree distributions, it seems almost obvious to look for a
deeper connection between networks and nonextensive statistical physics.
The purpose of this work is to show that various model types can be unified into a single dynamic 
network-formation model, characterized by
a reasonably small number of parameters. Within this parameter space, all
networks seem to be compatible with $q$-exponential degree
distributions.

\section{Model}

The following model is  a unification and generalization of the
models presented in \cite{tsallis_sm,thutsall}. The model in
\cite{tsallis_sm} captures preferential growing aspects of networks
embedded into a {\it metric space}, while \cite{thutsall} introduces
a static, selforganizing model with a sensitivity to an {\it
internal metric} (chemical distance, Diekstra distance). The
rewiring scheme there can be thought of having 
preferential attachment aspects in one of its limits \cite{sneppen}
(see below), but has none in the other limit.

\subsection{Network model}

The network evolves in time as described in
\cite{tsallis_sm}:  At $t=1$, the first node ($i=1$) is placed at
some arbitrary position in a metric space. The next node is placed
isotropically on a sphere (in that space) of radius $r$, which is drawn from a
distribution $P_G(r) \propto 1/r^{\alpha_G}$ ($\alpha_G>0$, $G$
stands for {\it growth}. To avoid problems with the singularity,
we impose a cutoff at $r_{\rm min}=1$. The second
node is linked to the first. The third node is placed again
isotropically on a sphere with random radius $r \in P_G$, however
the center of the sphere is now the barycenter of all the pre-existing
nodes. From the third added node on, there is an ambiguity where the
newly positioned node should link to. We choose a generalized
preferential attachment process, meaning that the probability that
the newly created node $i$ attaches to a previously existing node
$j$ is proportional to the degree $k_j$ of the existing node $j$,
and on the metric (Euclidean) distance between $i$ and $j$, denoted
by $r_{ij}$. In particular the linking probability is
\begin{equation}
p^A_{ij}= \frac{k_j / r^{\alpha_A}_{ij} }{ \sum_{j=1}^{N(t)-1}   k_j
/ r^{\alpha_A}_{ij}    } \quad ,
\label{PA}
\end{equation}
where $N(t)$ is the number of nodes at time $t$. It is not necessary
that at each time step only one node is entering the system, so we
immediately generalize that a number of $\bar n$ nodes are produced
and linked to the existing network with $\bar l$ links per time step.
Note that $\bar n$ and $\bar l$ can also be random numbers from an
arbitrary distribution.
For simplicity and clarity we fix $\bar n=1$ and $\bar l=1$.

After every $\lambda$ timesteps, a different action takes place on
the network. At this timestep the network does not grow but a pair
of nodes, say $i$ and $j$, merge to form one single node
\cite{sneppen}. This node keeps the name of one of the original
nodes, say for example $i$. This node now gains all the links of the
other node $j$, resulting in a change of degree for node $i$
according to
\begin{eqnarray}
k_i &\to& k_i + k_j -N_{\rm common}  \quad , \quad {\rm
\,\,\,\,\,\,\,\,\,\, if \,\,} (i,j){\rm \,\,are \,\, not \,\, first
\,\, neighbors}
\nonumber \\
k_i &\to& k_i + k_j -N_{\rm common}-2\quad , \quad {\rm if \,\,}
(i,j){\rm \,\,are \,\, first \,\, neighbors} \label{update}
\end{eqnarray}
where $N_{\rm common}$ is the number of nodes, which shared links to
both of $i$ and $j$ before the merger. In the case that $i$ and $j$
were first neighbors before the merger, i.e., they had been
previously linked, the removal of this link will be taken care of by
the term $- 2$ in Eq. (\ref{update}). The probability that two nodes
$i$ and $j$ merge can be made distance dependent, as before. In
particular  to stay close to the model in \cite{thutsall}, we chose
the following procedure. We randomly choose node $i$ with
probability $\propto 1/N(t)$ and then choose the merging partner $j$
with  probability
\begin{equation}
p^M_{ij}= \frac{d_{ij}^{-\alpha_M} }{\sum_j d_{ij}^{-\alpha_M}}
\;\;\;\;(\alpha_M \ge 0) \quad ,
\end{equation}
where $d_{ij}$ is the shortest distance (path) on the network
connecting nodes $i$ and $j$; Obviously, tuning $\alpha_M$ from $0$
toward large values, switches the model from the case where $j$ is
picked fully at random ($\propto 1/N(t)$), to a case where only
nearest neighbors of $i$  will have a nonnegligible chance to get chosen
for the merger. Note that the number of nodes is reduced by one at
that point.
To keep the number of nodes constant at this timestep, a new
node is introduced and linked with $\bar l$ of the existing nodes
with probability given in Eq. (\ref{PA}).

This concludes the model. Summing up, the relevant model parameters,
we have the merging exponent $\alpha_M$, the attachment exponent $\alpha_A$,
controlling the sensitivity of 'distance' in the network, and the
relative rate of merging and growing, $\lambda$. The parameters,
$\alpha_G$, $\bar n$, $\bar l$, and $r_{\rm min}$ 
have been found to play no major role in the model.

\begin{figure}[t]
 \begin{center}
 \begin{tabular}{cc}
 \includegraphics[height=60mm]{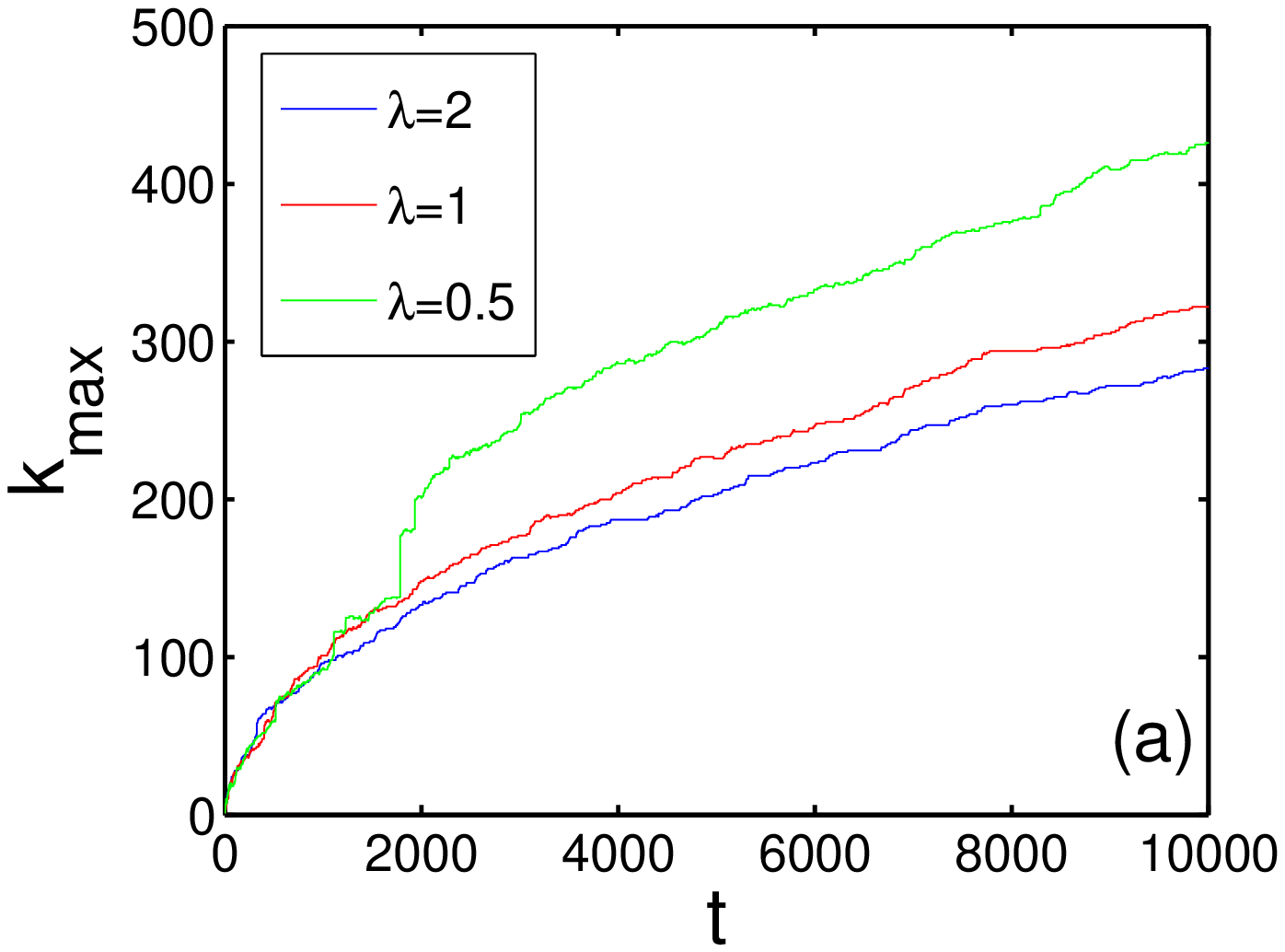}
 \includegraphics[height=60mm]{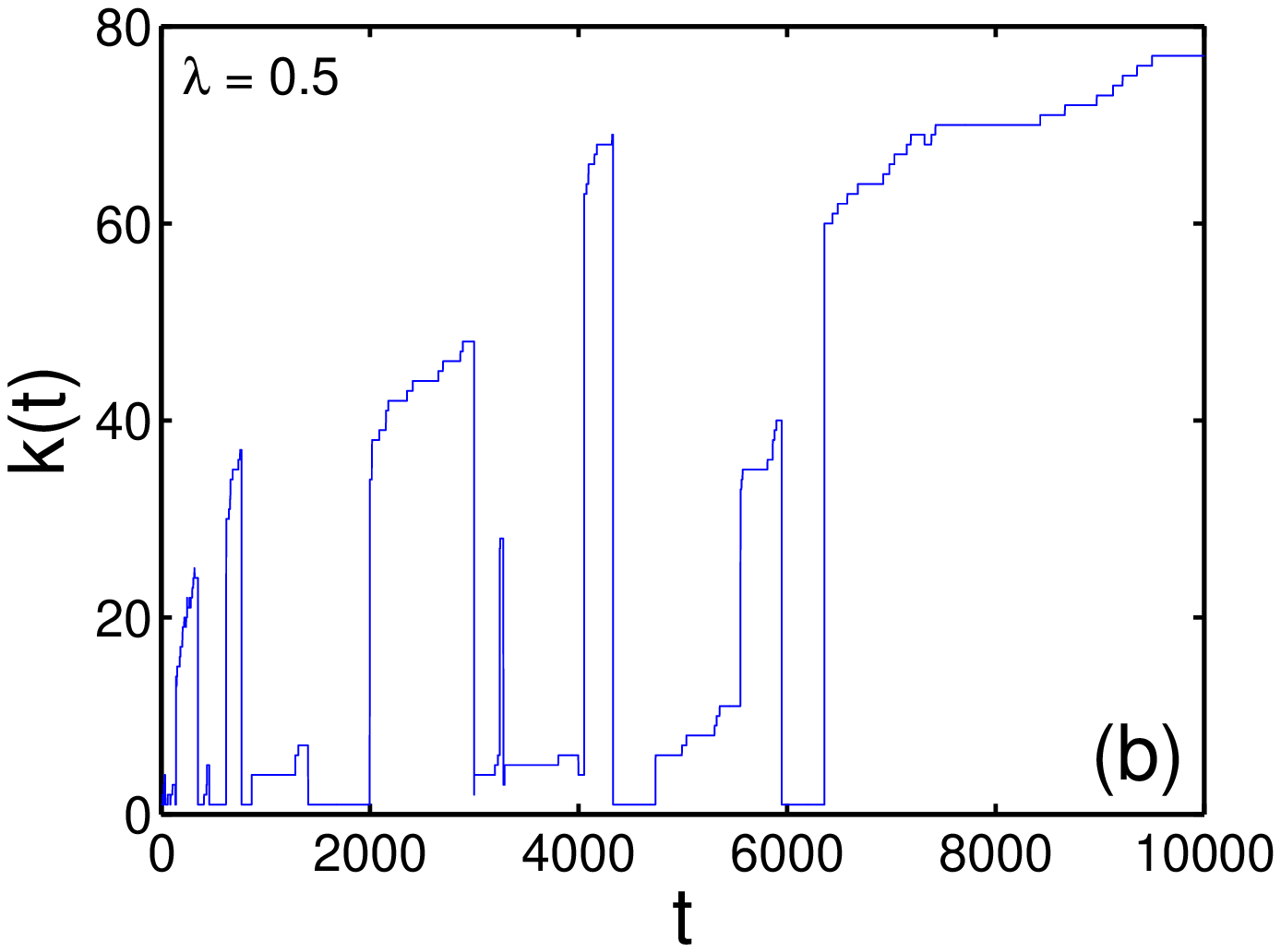}
 \end{tabular}
 \end{center}
 \caption{Time evolution of the degree of the best connected node (a) and of a
 randomly chosen node (b) for the parameters, $N = 10000$, $\alpha_A = 0 $, $\alpha_M =0$.}
 \label{Z}
\end{figure}

\begin{figure}[t]
 \begin{center}
 \begin{tabular}{cr}
 \includegraphics[height=80mm]{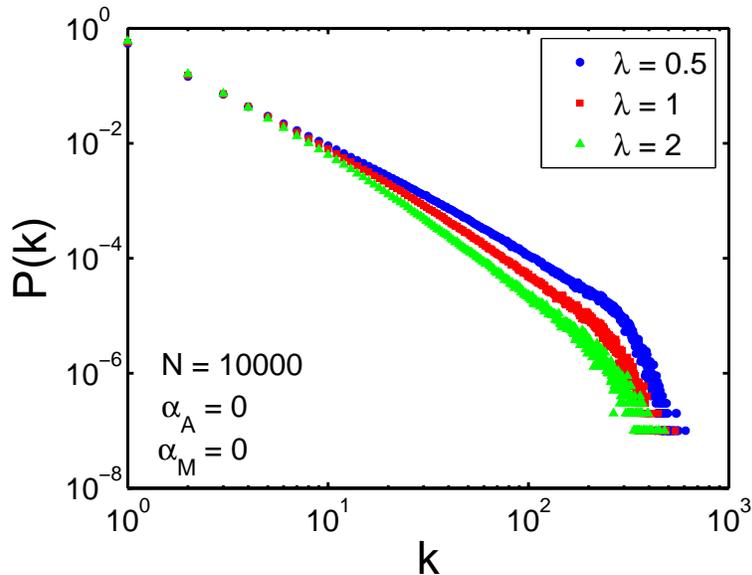}\\
 \end{tabular}
 \end{center}
 \caption{Degree distribution $P(k)$ (un-normalized) for $N = 10000$,  $\alpha_A=0$, $\alpha_M=0$ and various
 values of $\lambda$.} 
 \label{degr1}
 \end{figure}

We simulate this model and record the degrees $k_i$,
the clustering coefficients $c_i$ (defined below), and the nearest
neighbor-connectivity $k^{nn}_i$, for all individual nodes $i$.
From these values we derive distribution functions (as a function of $k$). In Fig. \ref{degr1}
typical degree distributions are shown for three typical values of $\lambda$. Obviously, 
the distribution is dominated by a power-law decay (see details of the functional form below) 
ending in an exponential finite size cut-off for large $k$.
 
The clustering coefficient of node $i$,  $c_i$ is defined by
\begin{equation}
 c_i=\frac{2e_i}{k_i(k_i-1)} \quad  ,
\end{equation}
with $e_i$ being the number of triangles node $i$ is part of. $c(k)$ is
obtained by averaging over all $c_i$ with a fixed $k$. It has been
noted that $c(k)$ contains information about
hierarchies present in networks \cite{ravas}. For Erd\"os-R\'enyi (ER)
networks \cite{ER}, as well as for pure preferential attachment
algorithms without the possibility of rewiring,  the clustering
coefficient $c(k)$ vs. degree is  flat. The global clustering
coefficient is the average over all nodes $C=\langle c_i\rangle_i$.
A large global clustering coefficient is often used for
identification of small-world structure \cite{watts}. The average
nearest-neighbor connectivity (of the neighbors) of node $i$ is
\begin{equation}
 k^{nn}_i=    \frac{1}{k_i}  \sum_{j \, {\rm neighbor \, of} \, i}  k_j
   \quad .
\end{equation}
When plotted as a function of $k$, $k^{nn}(k)$ is a measure to
assess the assortativity of networks. A rising function means
assortativity, which is the tendency for well connected nodes to
link to other well connected ones, while a declining function
signals disassortative structure.

\subsection{Particular instances of the model}
Depending on the variables of the model, known networks result as
natural limits.

\subsubsection{Soares et al. limit}
For the $\lim \lambda\to \infty$ we have no merging, and $\alpha_M$
is an irrelevant parameter. The model corresponding to this limit
has been proposed and studied in \cite{tsallis_sm}.

\subsubsection{Albert-Barabasi limit}
The $\lim \lambda\to \infty$ and $\lim \alpha_A\to0$, gets rid of
the metric in the Soares et al. model and recovers the original
Albert-Barabasi preferential attachment model.

\subsubsection{Kim et al. limits}
The limit $\lim \lambda\to 0$ allows no preferential growing of the
network. If at each timestep after every merger a new node is linked
randomly with $\bar l$ links to the network, the model reported in
\cite{thutsall} is recovered. The $\lim \lambda\to 0$ model with $\lim
\alpha_M\to0$ ($\lim \alpha_M\to \infty$) recovers the {\it random} case ({\it neighbor} case) in \cite{sneppen}.

\section{Nonextensive characterization of complex networks}

There has been a convincing body of evidence that, for a large class
of networks, (normalized) degree distributions can be fit by
$q$-exponentials,
\begin{equation}
  P(k)= e_{q}^{-(k-1)/\kappa} \;\;\;\;(k =1,2,3,4,...)\quad ,
  \label{ans}
\end{equation}
where the $q$-exponential function is defined in Eq.
(\ref{qexp}), with $q \ge 1$, and $\kappa>0$ some characteristic number of links.
A convenient procedure to perform a two-parameter fit of this kind
is to take the {\it $q$-logarithm} of the distribution $P$, defined
by $Z_{q}(k) \equiv \ln_{q} P(k) \equiv  \frac{[P(k)]^{1-q}-1}{1-q}$. This is done for a series of different
values of $q$. The function $Z_{q}(k)$ which can be best fit with a
straight line determines the value of $q$, the slope being $-\kappa$.
\begin{figure}[t]
 \begin{center}
 \begin{tabular}{c}
 \includegraphics[height=65mm]{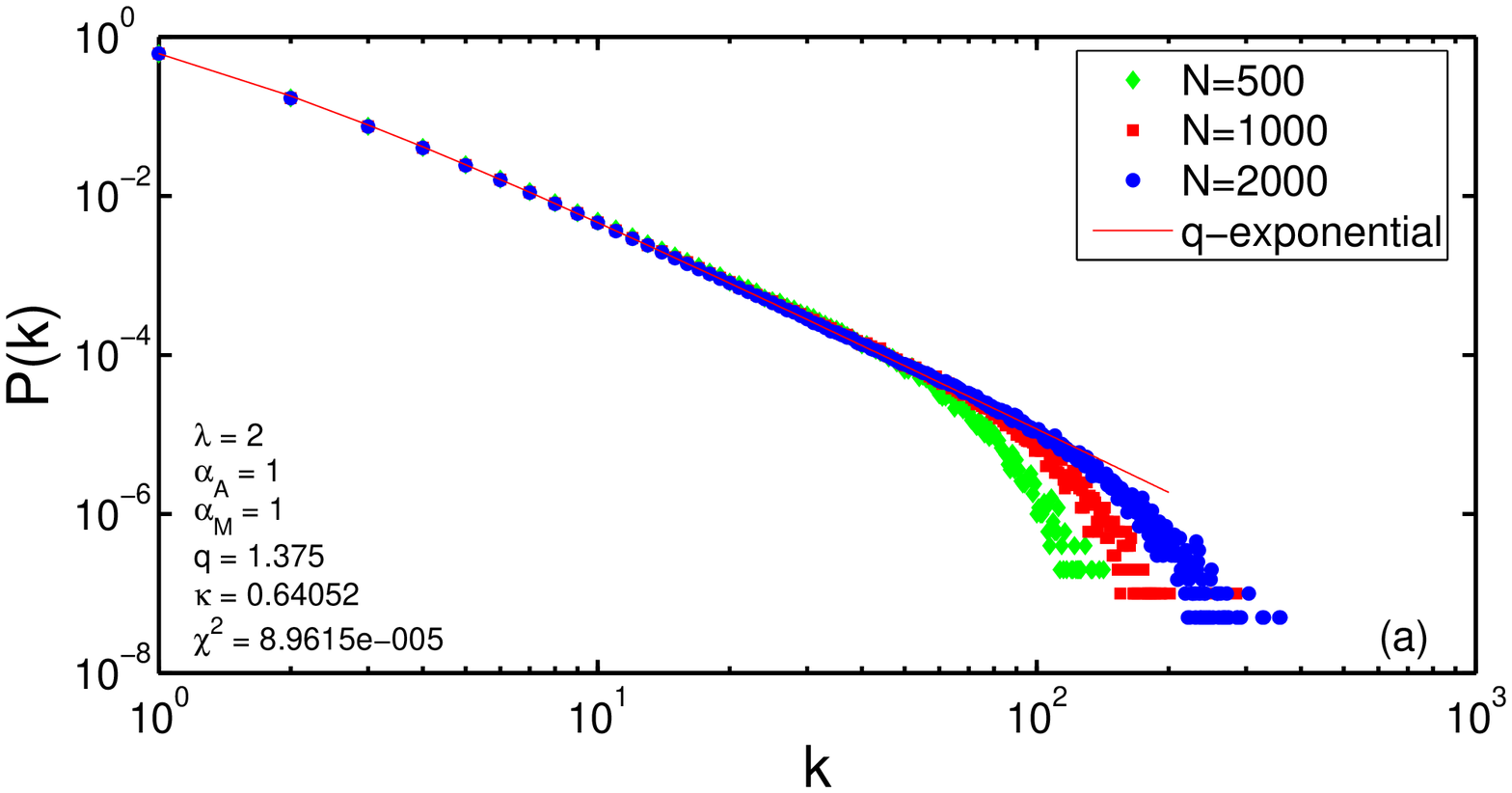}\\
 \includegraphics[height=65mm]{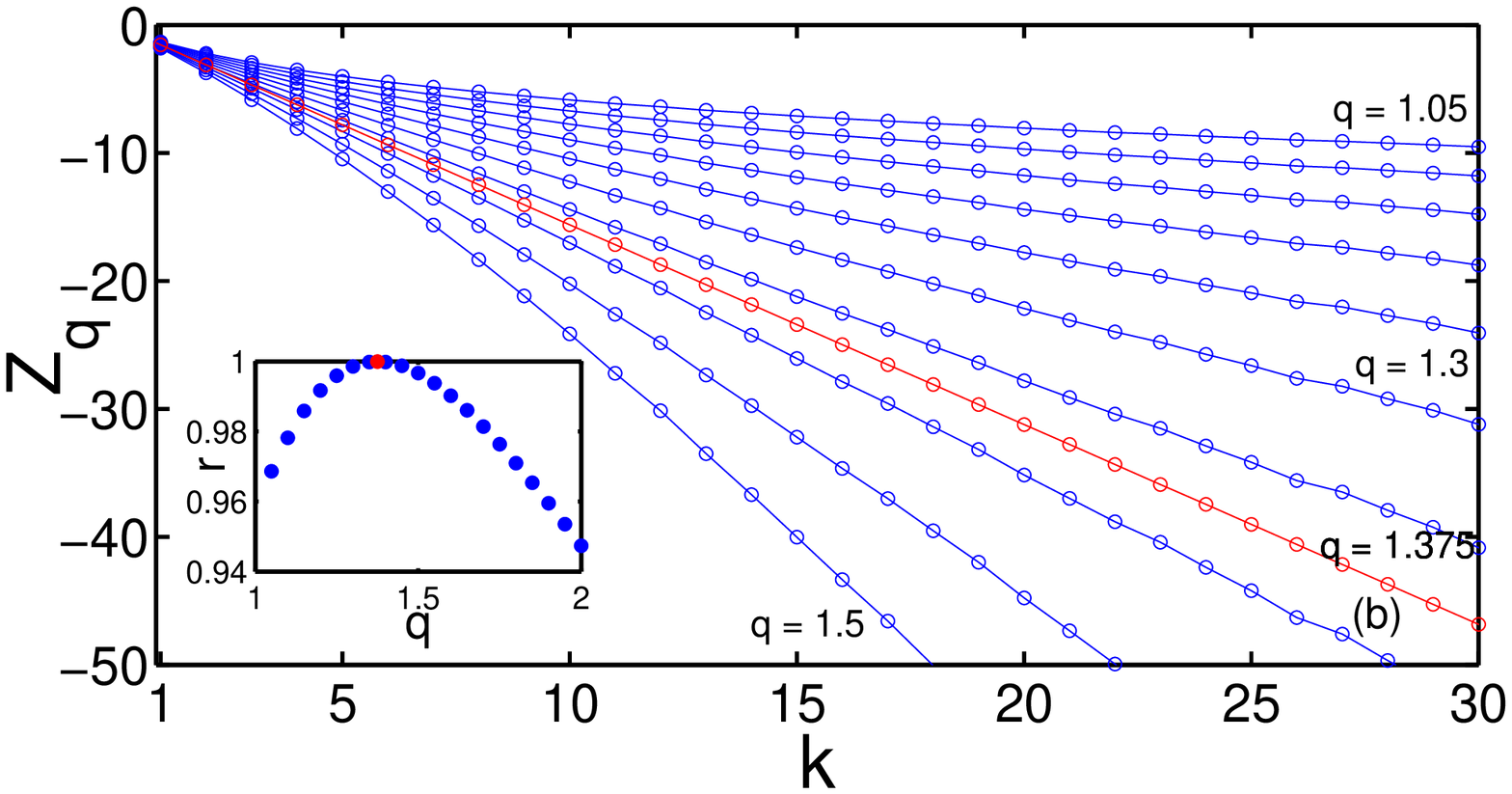}\\
 \end{tabular}
 \end{center}
 \caption{ 
 (a) $P(k)$ for $\lambda=2, \alpha_A=1, \alpha_M=1$, and
 various system sizes (symbols). The line is the $q$-exponential fit for $N=2000$.
 (b) $q$-logarithm of the (normalized) $P(k)$ from (a).  The line associated with 
 $q=1.375$ corresponds to an optimal linear fit, i.e. a maximum of the correlation 
 coefficient (inset) of a straight line with $Z_q$. The quality of the fit in (a) is 
 given by a standard $\chi^2$ statistics.}
 \label{Pc-scale-lamda}
\end{figure}

In Fig. \ref{Pc-scale-lamda} we show the degree distribution for several 
system sizes together with the $q$-logarithm $Z_q(k)$, from which an 
optimum $q$ and $\kappa$ can be obtained.  We conclude that, with good precision,  
the {\it Ansatz}  in Eq. (\ref{ans}) for the degree distribution,  when seen as a null hypothesis, 
can not be rejected on the basis of a $\chi^2$ statistics for any reasonable significance level,
for the system sizes studied.

For actual curve fitting, it is often more convenient to use the {\it cumulative}
distributions, which can be  parametrized by 
\begin{equation}
P(\geq k)= e_{q_c}^{-(k-1)/\kappa_c} \;\;\;\;(k =1,2,3,4,...)\quad .
\label{pgeqk}
\end{equation}
On the other hand the corresponding cumulative distribution $P(\geq k)$ is given by
(we switch to integral notation for simplicity for a moment)
\begin{equation}
P(\geq k) \equiv 1 - \int_1^k dk^\prime \, P(k^\prime) = \left[1-\frac{1-q}{\kappa} (k-1) \right]^{\frac{2-q}{1-q}} \quad .
\end{equation}
By comparison of coefficients the cumulative parameters are given  by 
\begin{equation}
  q_c = \frac{1}{2-q} \quad {\rm and} \quad \kappa_c = \frac{\kappa }{2-q} \quad .
\end{equation}
Whenever we talk about $q$-values corresponding to a cumulative
distribution, we  use the notation $q_c$ and $\kappa_c$, where $c$ indicates {\it
cumulative}.

The remarkable quality of $q$-exponential fits to the degree distributions 
from the model, reveals a connection \cite{tsallis_sm} of
scale-free network dynamics to nonextensive statistical mechanics
\cite{tsallis88,gellmann}. To make the point more clear, consider the entropy
\begin{equation}
S_q \equiv \frac{1-\int_1^\infty dk \, [p(k)]^q}{q-1} \;\;\;\;\;\;\;
\Bigl[S_1=S_{BG} \equiv -\int_1^\infty dk \, p(k) \ln p(k)\Bigr] \,,
\end{equation}
where we assume $k$ as a continuous variable for simplicity. If we
extremize $S_q$ with the constraints \cite{TsallisMendesPlastino}
\begin{equation}
 \int_1^\infty dk \, p(k)=1 \quad
 {\rm and} \quad
 \frac{\int_1^\infty dk \, k \, [p(k)]^q}{ \int_1^\infty dk \,
 [p(k)]^q  } = K\,,
 \label{constr2}
\end{equation}
we obtain
\begin{equation}
p(k)= \frac{e_q^{-\beta (k-1)}}{\int_1^\infty dk^\prime  \,
e_q^{-\beta (k^\prime-1)}}=\beta(2-q) e_q^{-\beta (k-1)} \;\;\;\;(k \ge 1) \,,
\end{equation}
where $\beta$ is determined through Eq. (\ref{constr2}). Both
positivity of $p(k)$ and the normalization constraint
(\ref{constr2}) impose $q<2$. 

Let us mention that models do exist that can be handled analytically, and which 
exhibit precisely $q$-exponential degree distributions. Such is the case of 
\cite{albert00}. The degree distribution is there presented in the form 
$p(k) \propto 1/(k+k_0)^\gamma$. This form can be re-written as a $q$-exponential with 
$q=\frac{\gamma + 1}{\gamma}=\frac{2m(2-r)+1-p-r}{m(3-2r)+1-p-r}$, where $(m,p,r)$ are 
parameters of the particular model in \cite{albert00}.

\section{Results}

Realizing the above network model in  numerical simulations we
compute degree distributions, clustering coefficients, and neighbor
connectivity, for a scan over the relevant parameter space, spanned by 
$\lambda$, $\alpha_A$ and  $\alpha_M$. All following data were
obtained from averages over 100 identical network realizations with
a final $N(t^{\rm max})=1000$; for finite size checks we have
included runs with  $N(t^{\rm max})=500$ and $2000$. 
The reason for these relatively modest network sizes is that, at every timestep,  
all network distances have to evaluated. The remaining
parameters have been checked to be of marginal importance and  have
been fixed to $\alpha_G=1$, $\bar n=1$ and $\bar l =1$.

\begin{figure}[htb]
\begin{center}
\begin{tabular}{ccc}
{\Large $q_c$ } & {\Large $\kappa_c$ } & {\Large $\chi^2$ }\\
\includegraphics[height=40mm]{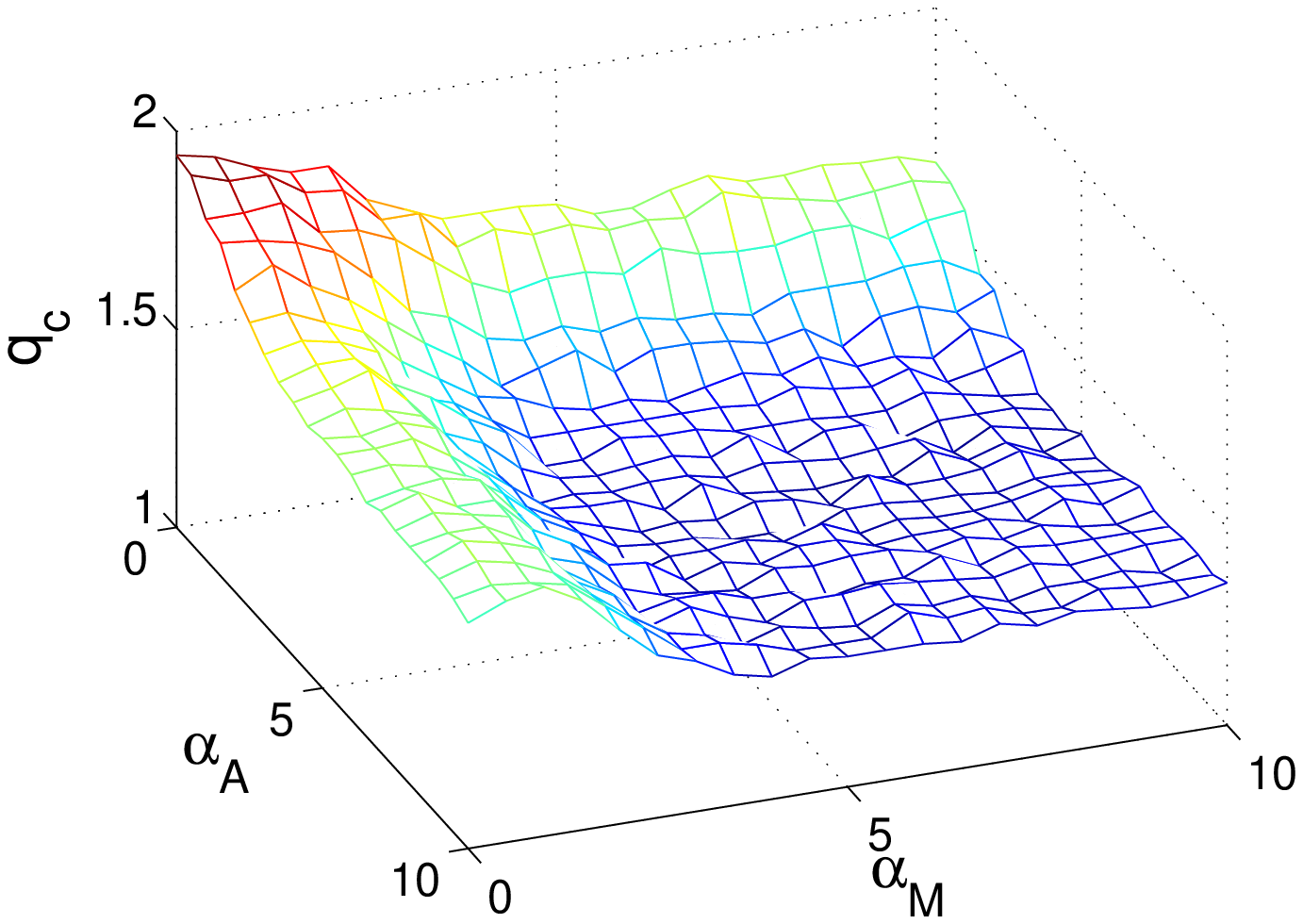} &
\includegraphics[height=40mm]{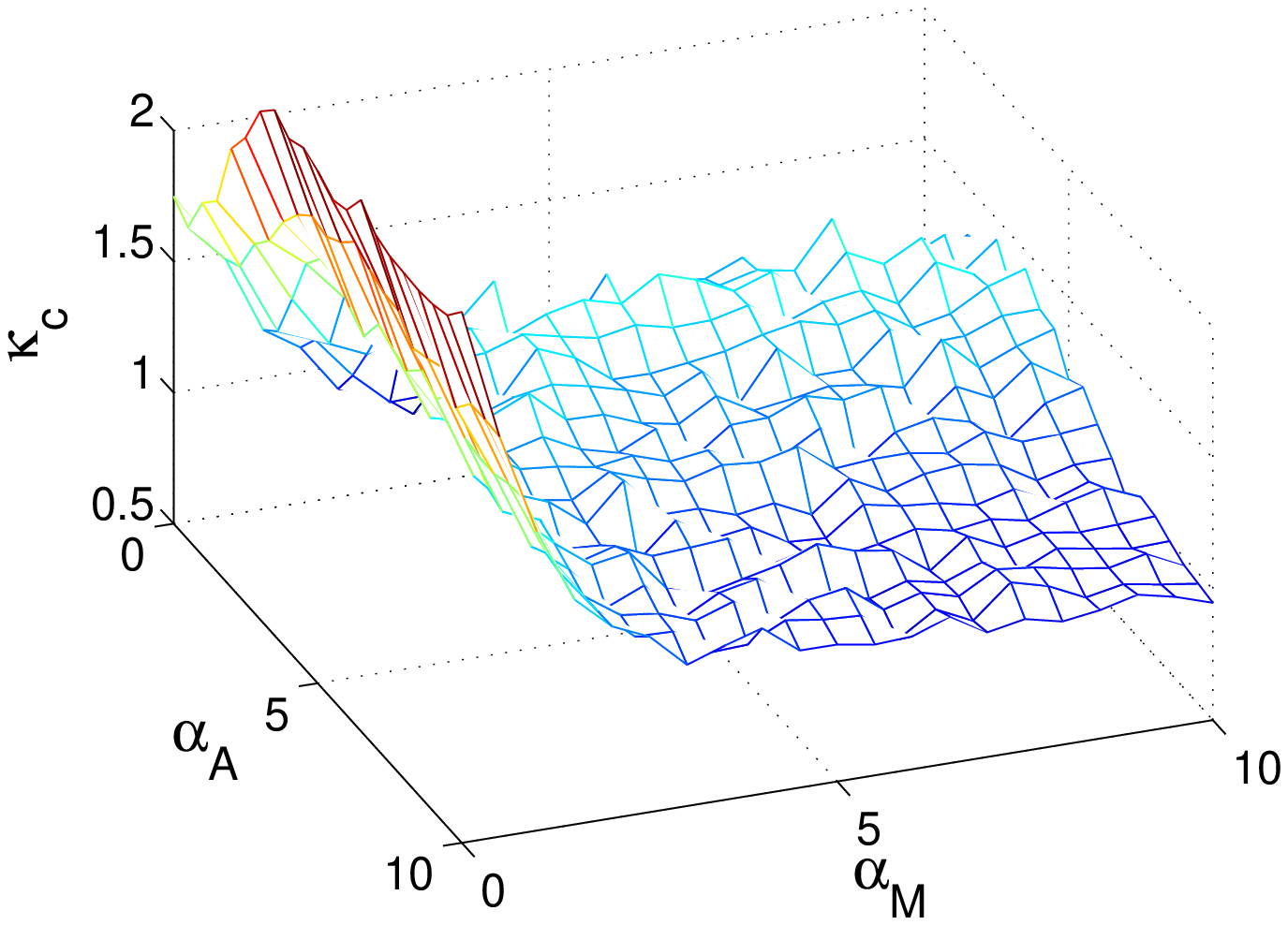} &
\includegraphics[height=40mm]{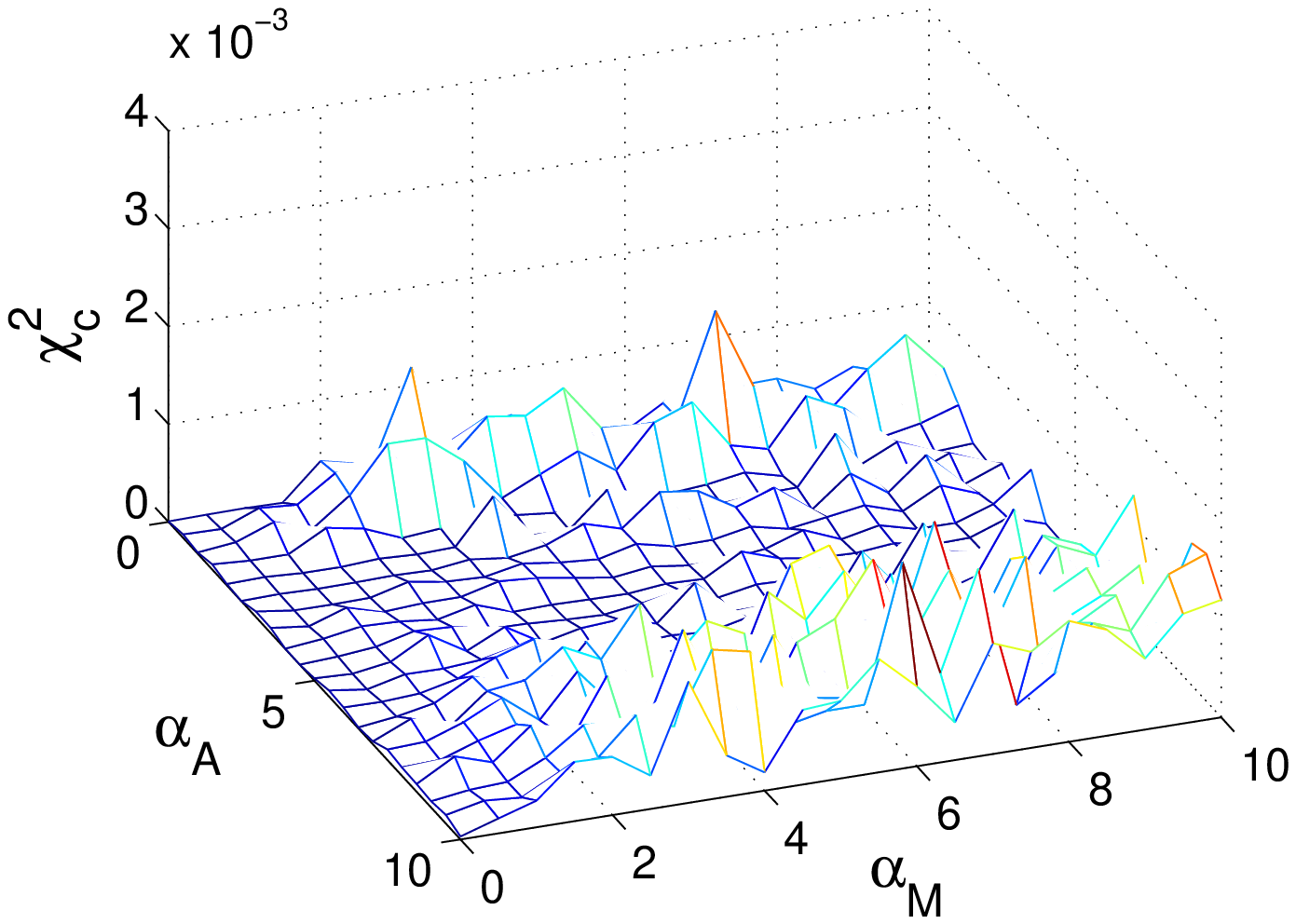} \\
\includegraphics[height=40mm]{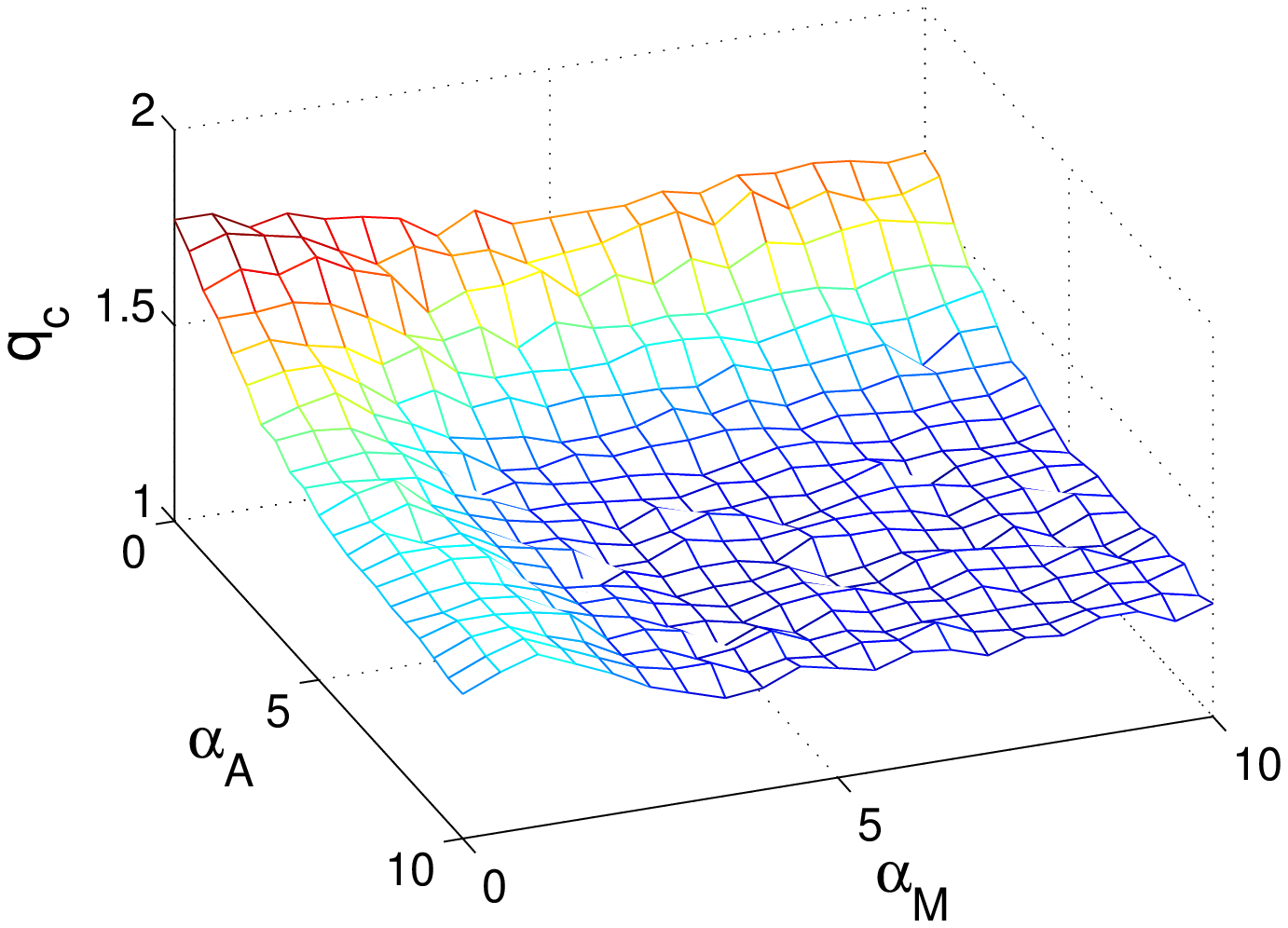} &
\includegraphics[height=40mm]{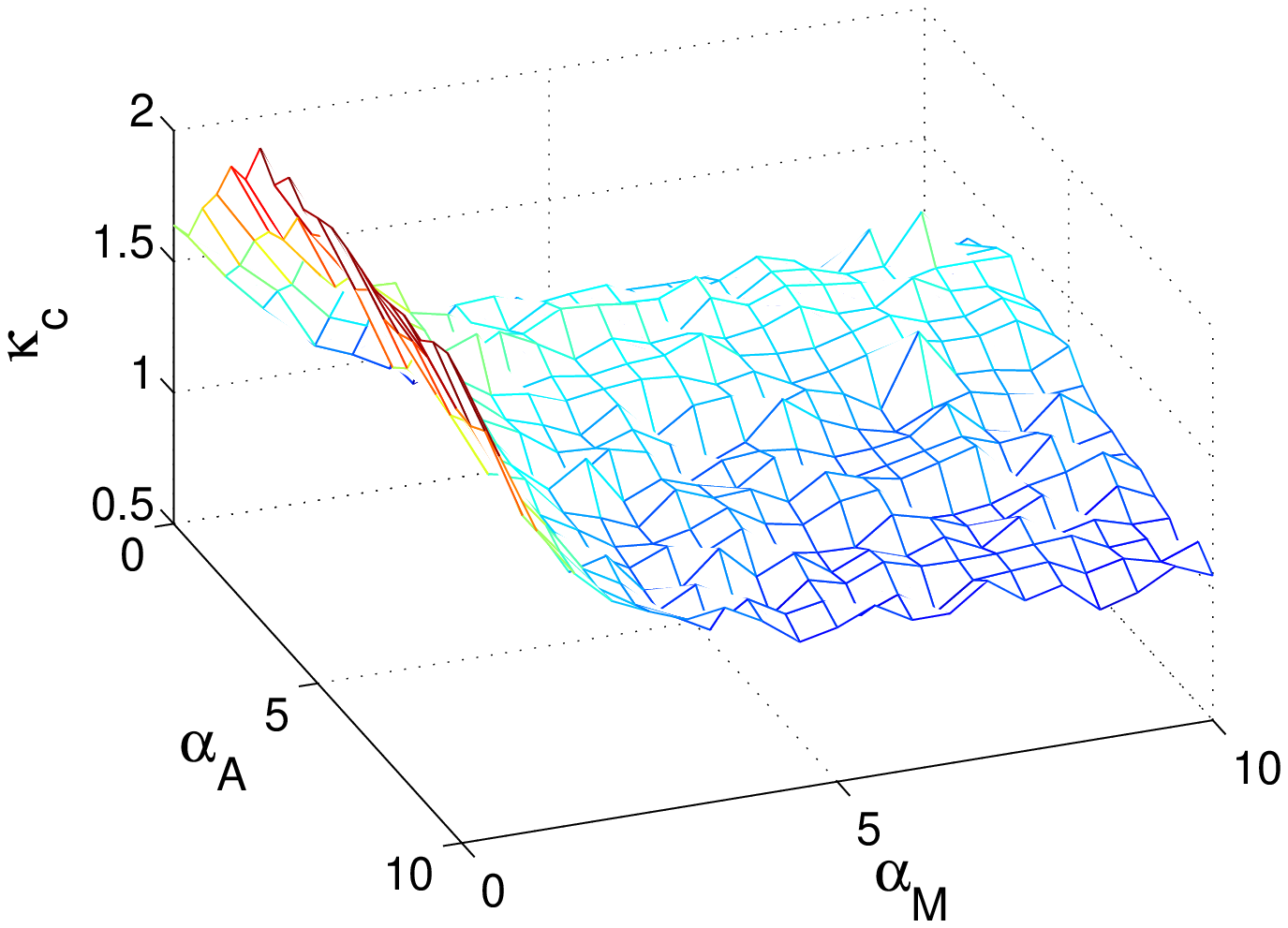} &
\includegraphics[height=40mm]{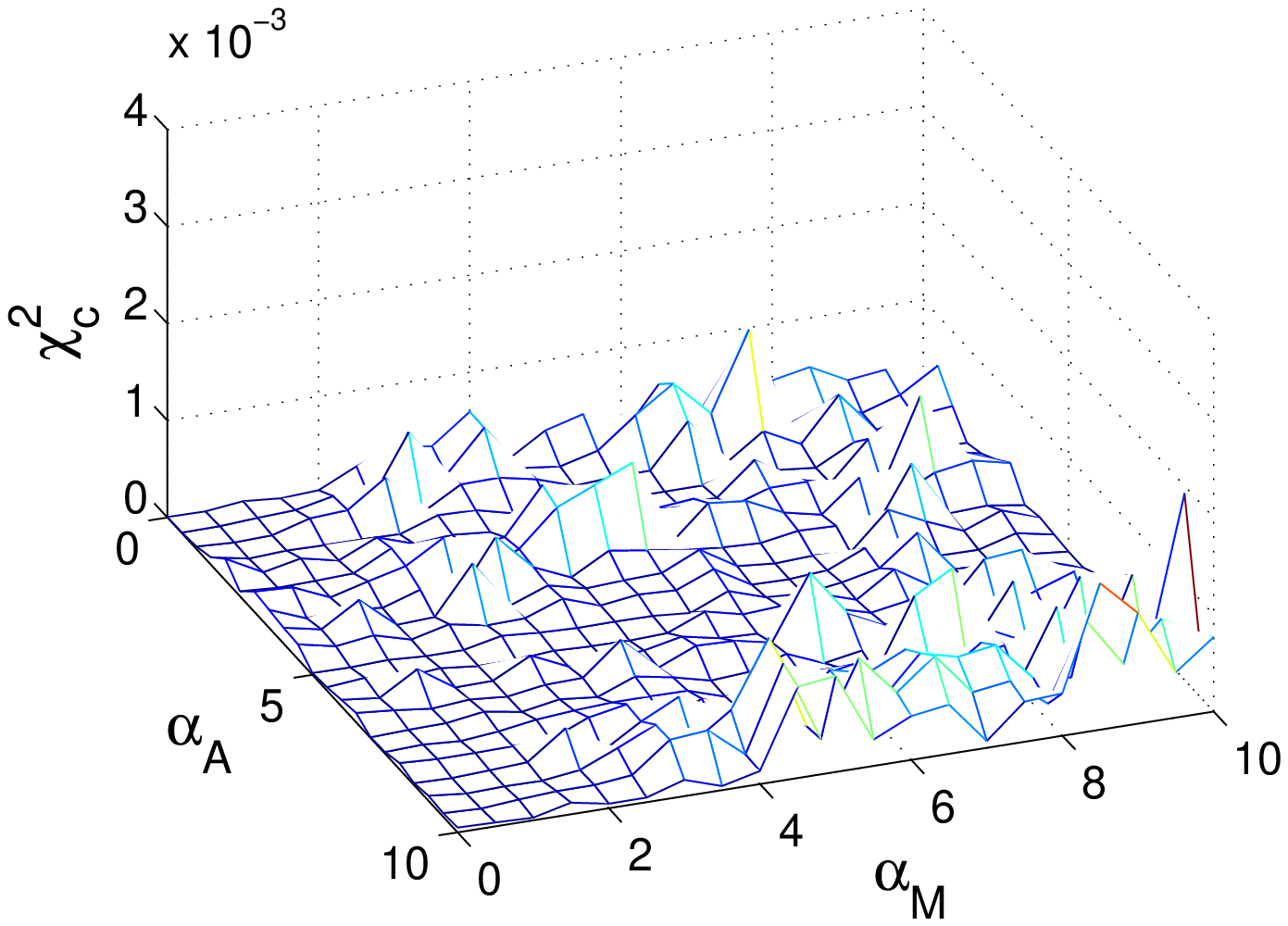} \\
\includegraphics[height=40mm]{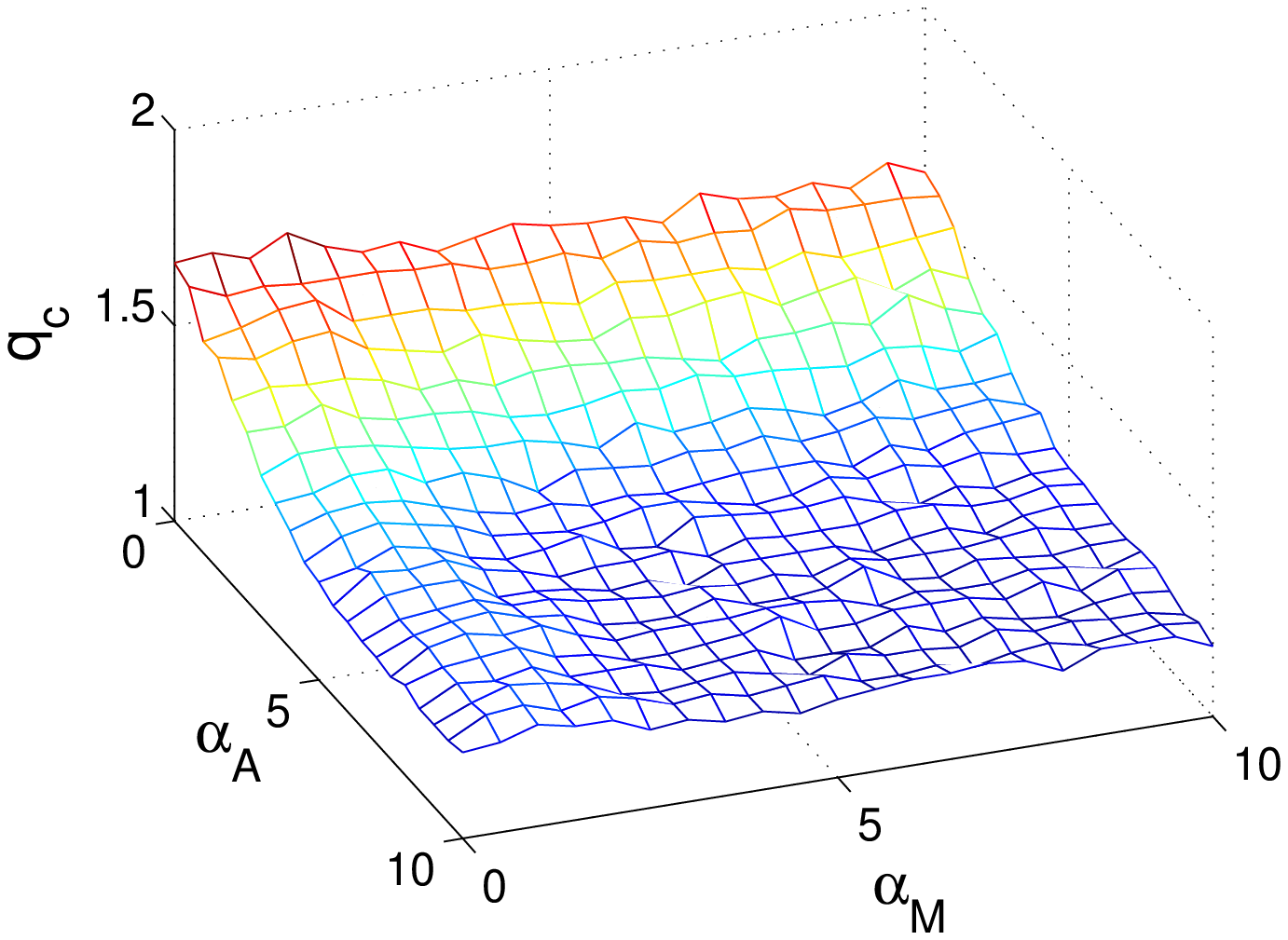} &
\includegraphics[height=40mm]{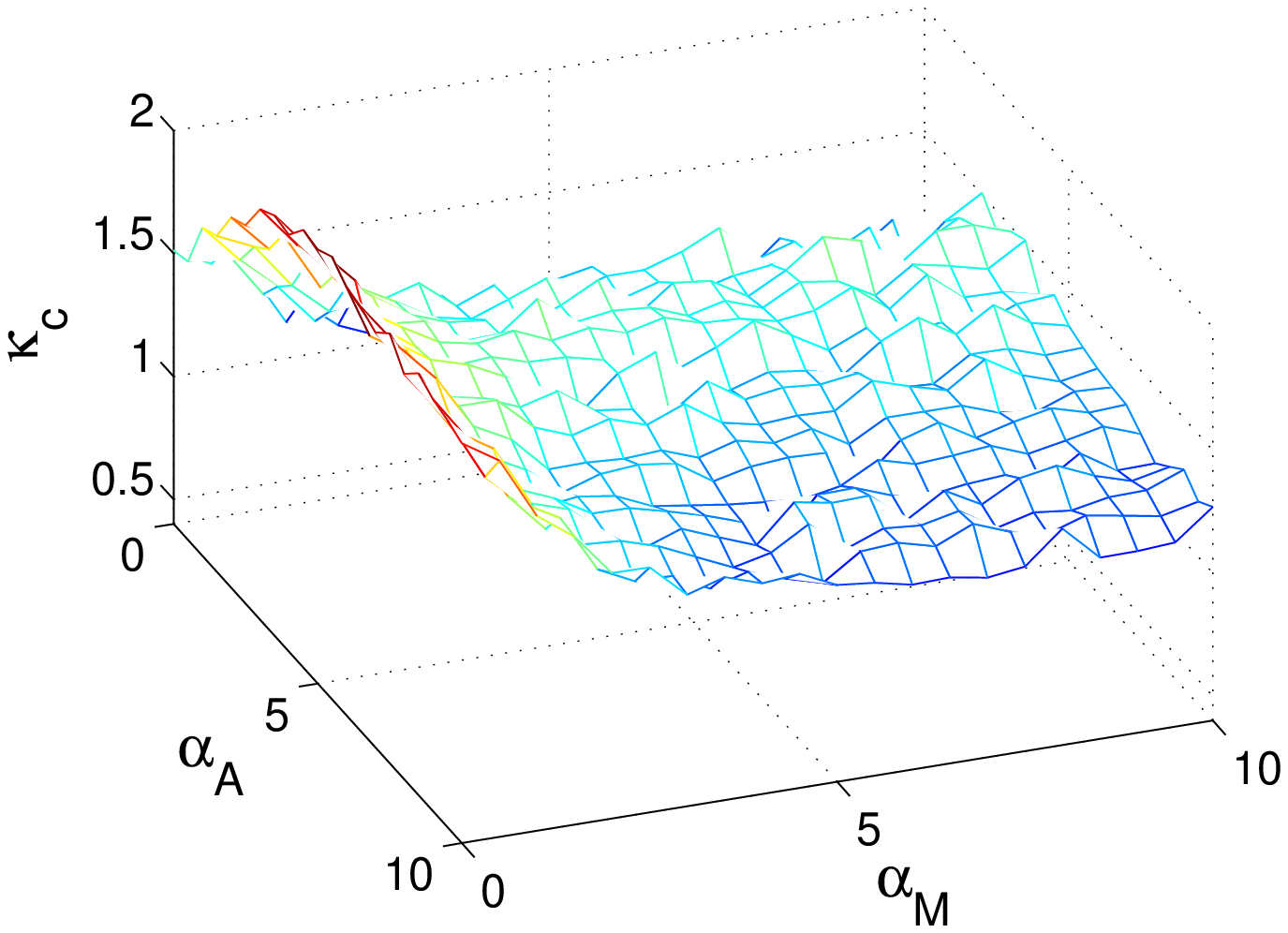} &
\includegraphics[height=40mm]{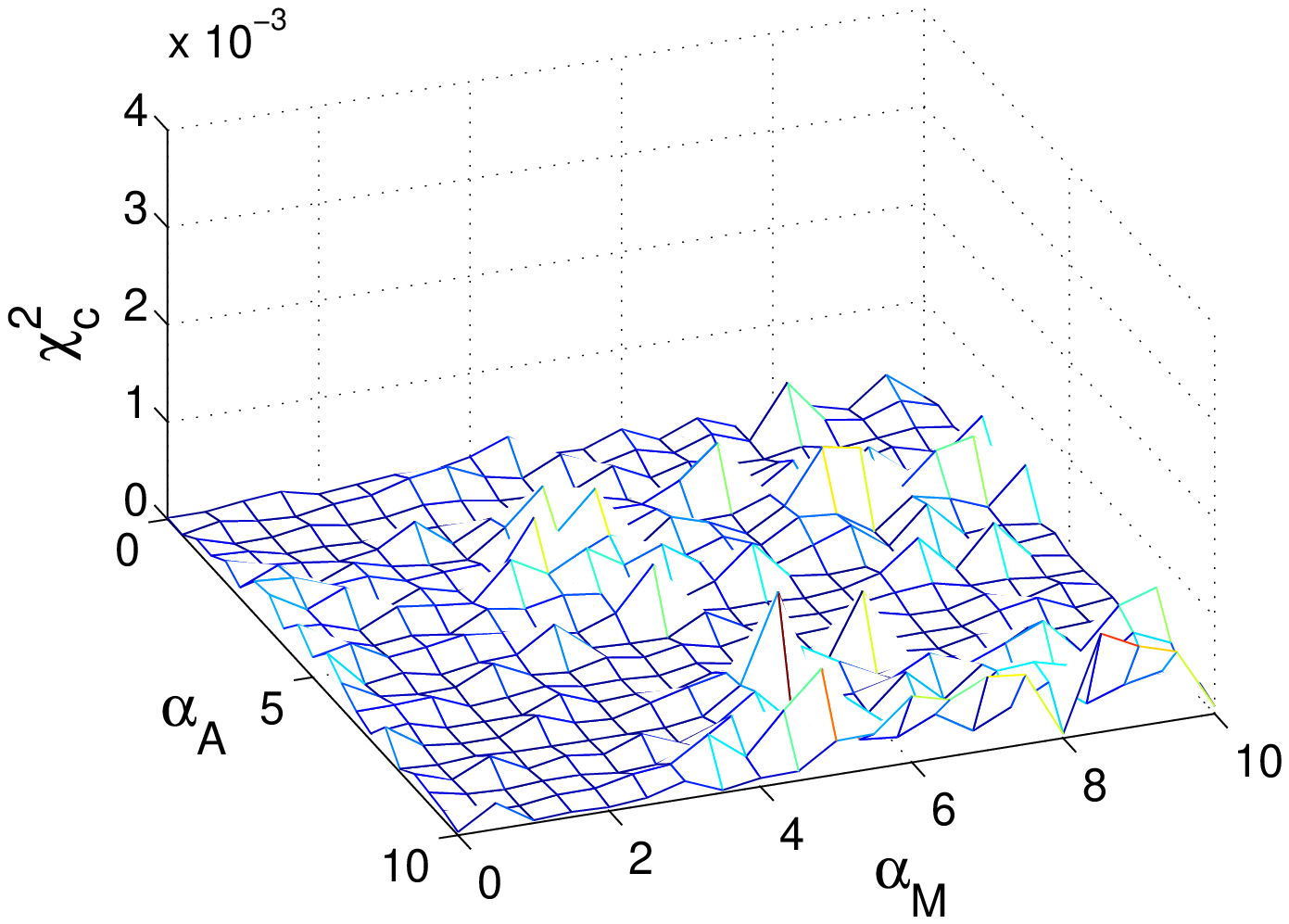} \\

\end{tabular}
\end{center}
\caption{$q_c$ and $\kappa_c$ values from $q$-exponential fits to the cumulative degree
distributions $P(>k)$ for $\alpha_G=1$, $N=1000$, and $\lambda=0.5$
(top), $\lambda=1$ (middle), $\lambda=2$ (bottom). The fit-quality is 
given by the $\chi^2$ value per degree of freedom.} 
 \label{qc-kc-alpha-alpha}
\end{figure}

The fitted values for the  nonextensivity index $q_c$ and the
characteristic degree $\kappa_c$ are shown in  Fig.
\ref{qc-kc-alpha-alpha} over the parameter space. From top to bottom
three values of $\lambda$ are shown. The $q_c$ index is declining in
all three parameters, $\alpha_A$, $\alpha_M$, and $\lambda$. It
eventually converges to a plateau in the $\alpha_A-\alpha_M$-plane.
The height of the plateau slowly decreases with higher $\lambda$,
but remains above $1$; $q_c=1$ corresponds to the exponential (ER) case. 
For  low $\alpha_M$ there is a maximum of $\kappa_c$ at about $\alpha_A \sim 3$;
For larger $\alpha_M$ a plateau is forming for all $\alpha_A$. This plateau 
remains constant as a function of $\lambda$.
The quality of the $q$-exponential fit is demonstrated by the $\chi^2$ test statistics 
per degree of freedom.

\begin{figure}[t]
 \begin{center}
 \begin{tabular}{cc}
 \includegraphics[height=60mm]{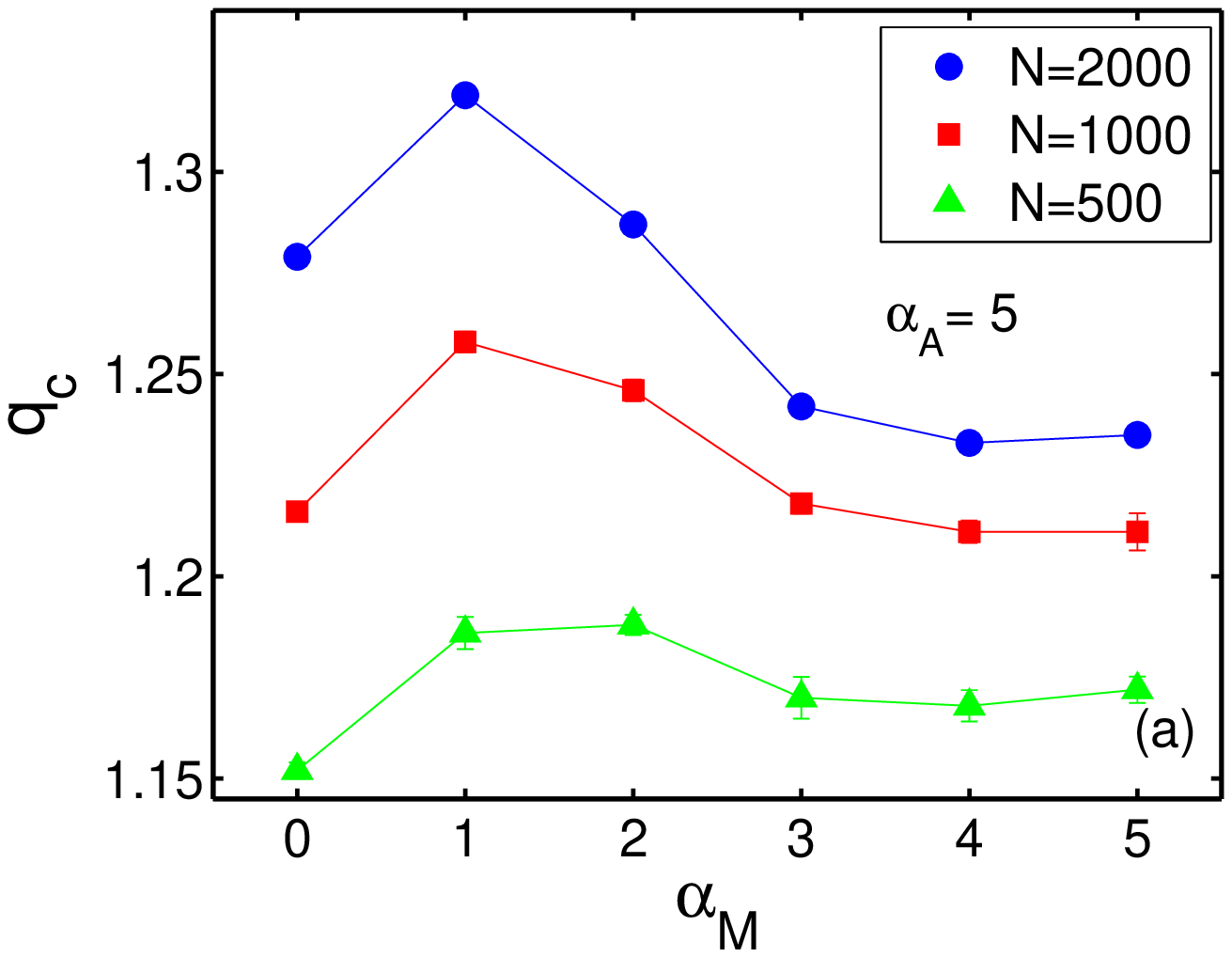}&
 \includegraphics[height=60mm]{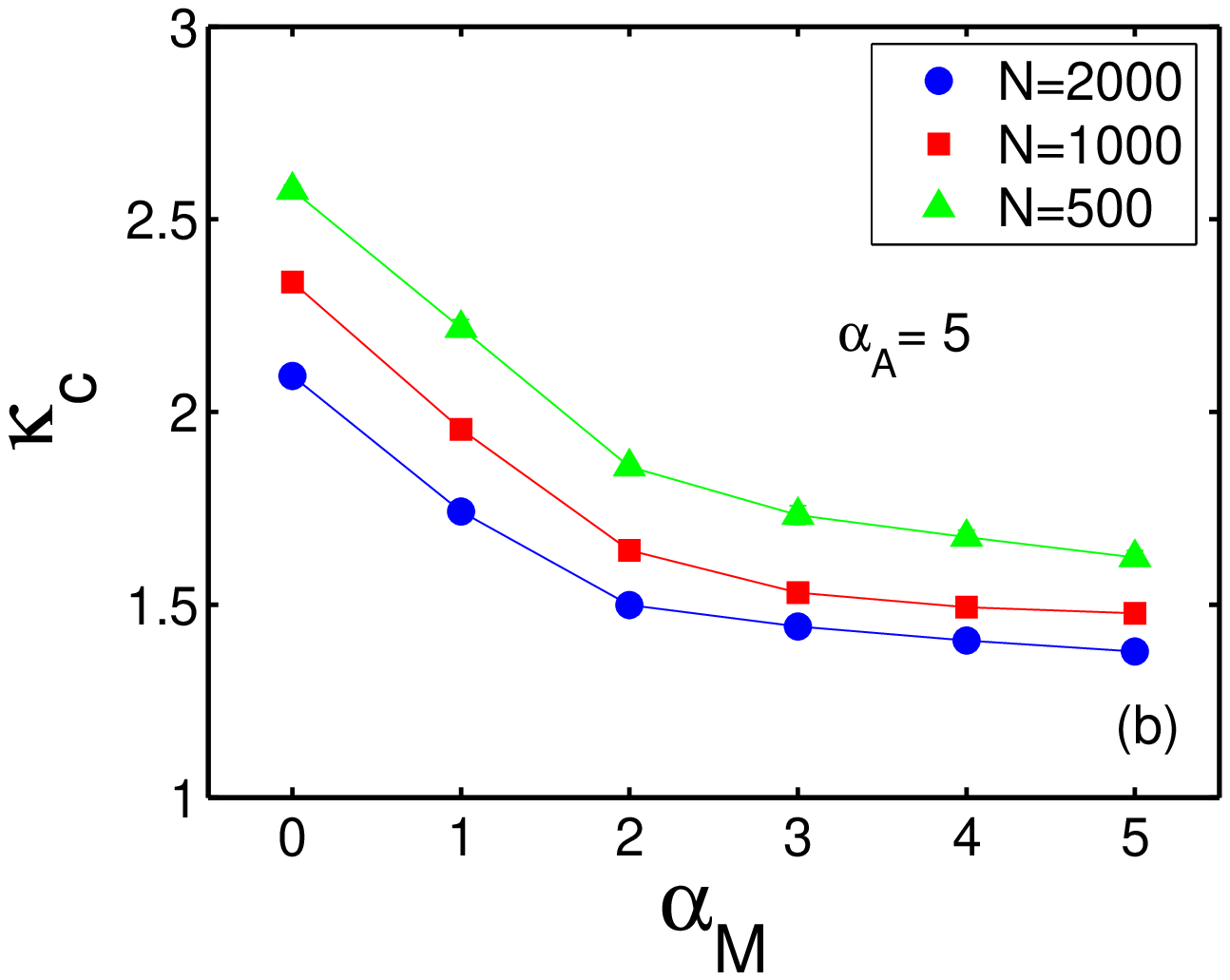}\\
 \includegraphics[height=60mm]{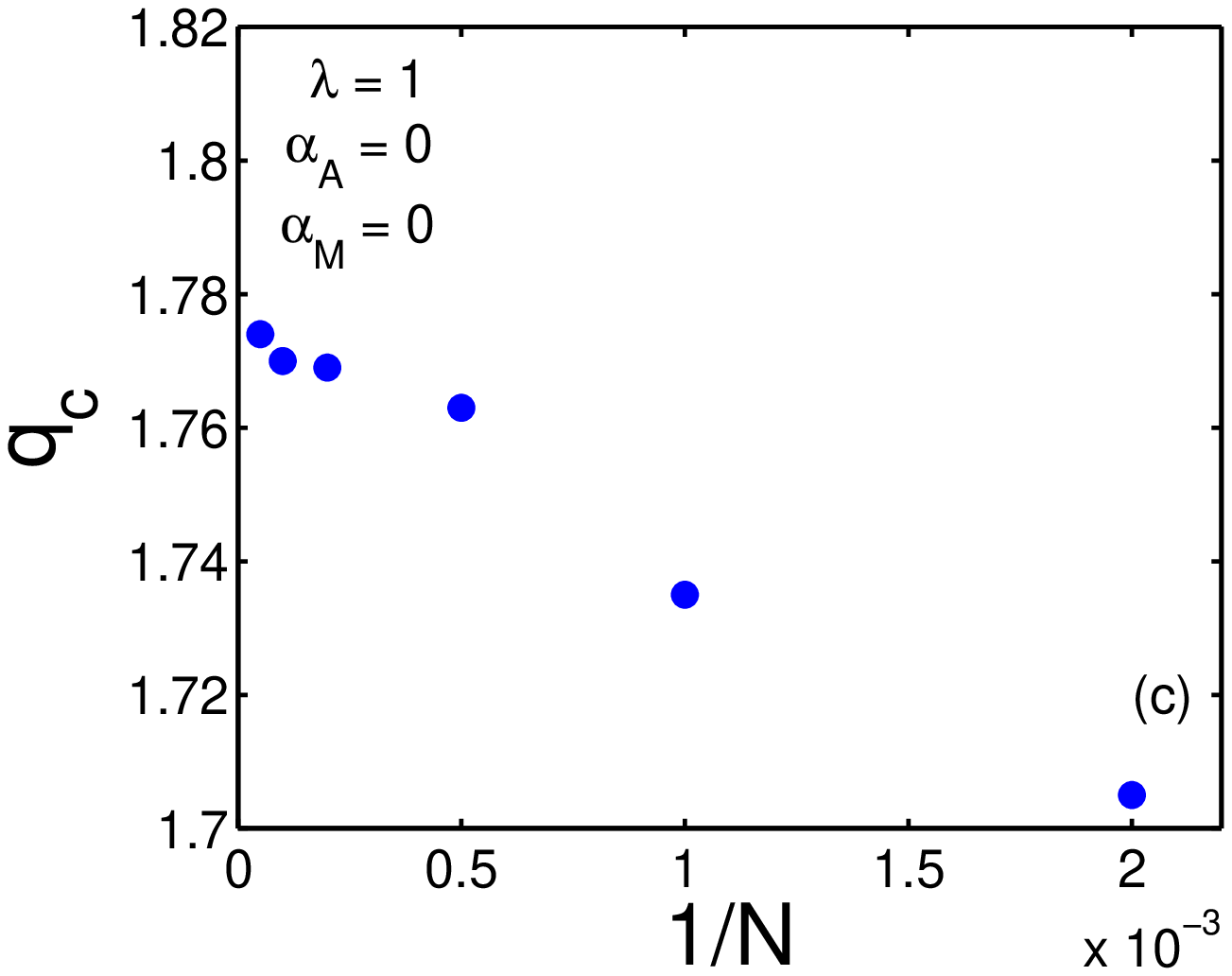}&
 \includegraphics[height=60mm]{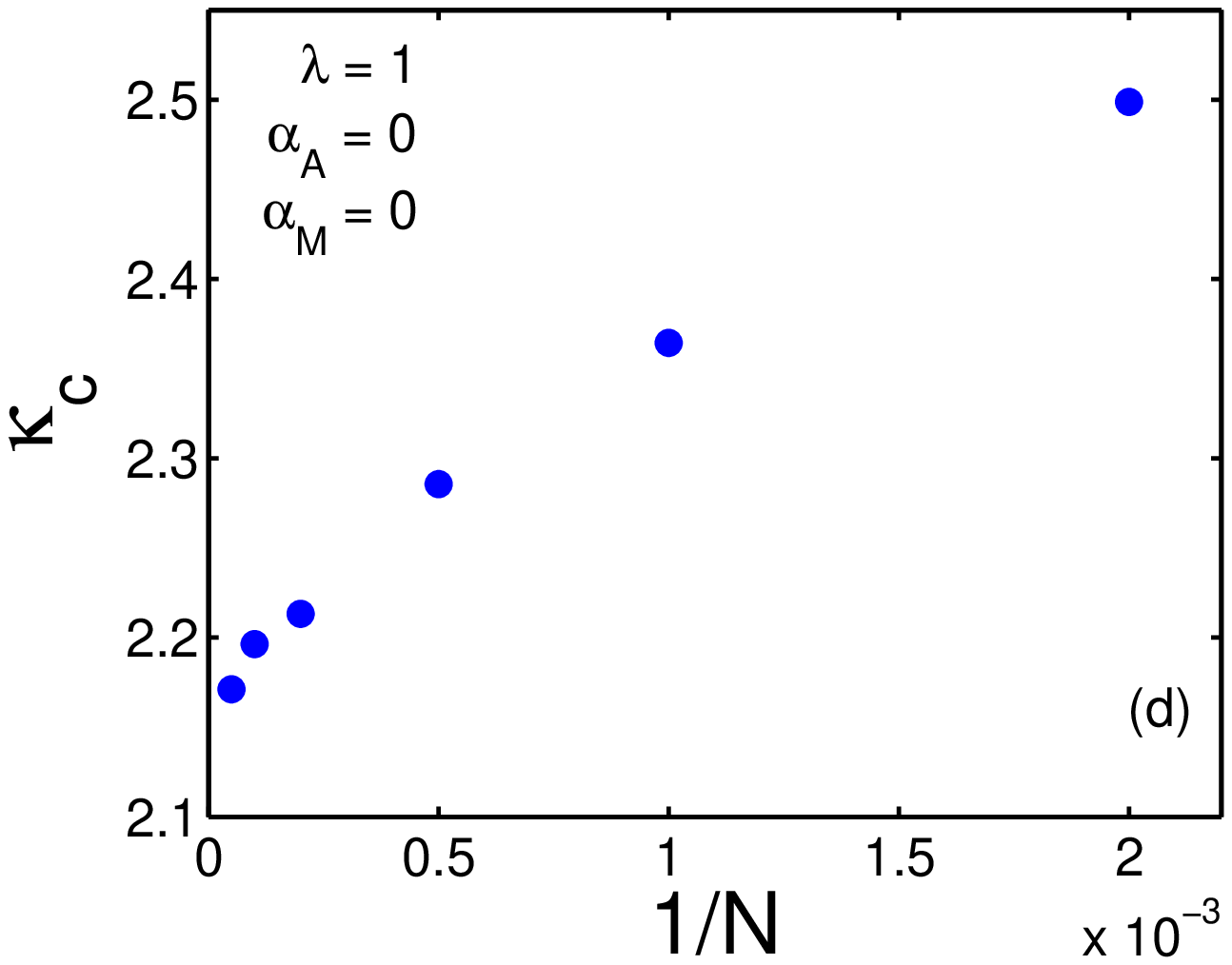}\\
 \end{tabular}
 \end{center}
 \caption{(a) $q_c$ values of 3 system sizes for
 $\lambda=2$, $\alpha_A=5$, and $\alpha_M$ ranging from 0 to 5. (b) same for  $\kappa_c$.
(c) and (d) show the same parameters as a function of network size $N$, 
for $\lambda=1$, $\alpha_A=\alpha_M=0$. For these parameters networks up to a size 
of $N=20000$ were possible. 
} 
\label{qscale-kscale}
\end{figure}

As in \cite{thutsall} we observe a finite size effect in the
data. In Fig. \ref{qscale-kscale} (a) we show the dependence
of the degree distribution parameters as a function of
$\alpha_M$ for different system sizes for a fixed $\alpha_A=5$, and 
$\lambda=2$. 
The fits  for $\kappa_c$ are shown in Fig. \ref{qscale-kscale} (b).

\begin{figure}[t]
 \begin{center}
 \begin{tabular}{cc}
 {\Large clustering} & {\Large neighbor conn.}\\
 \includegraphics[height=60mm]{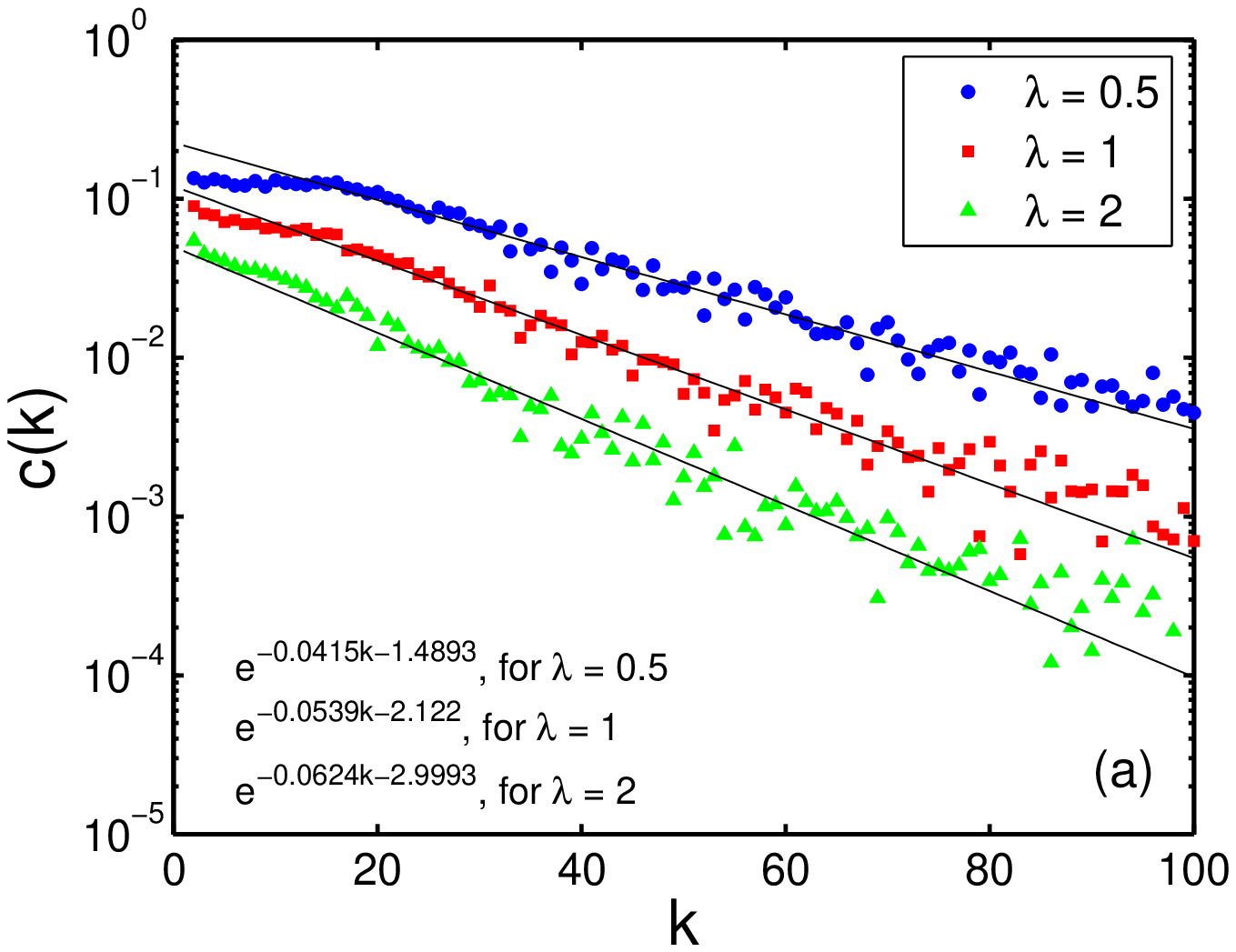}&
 \includegraphics[height=60mm]{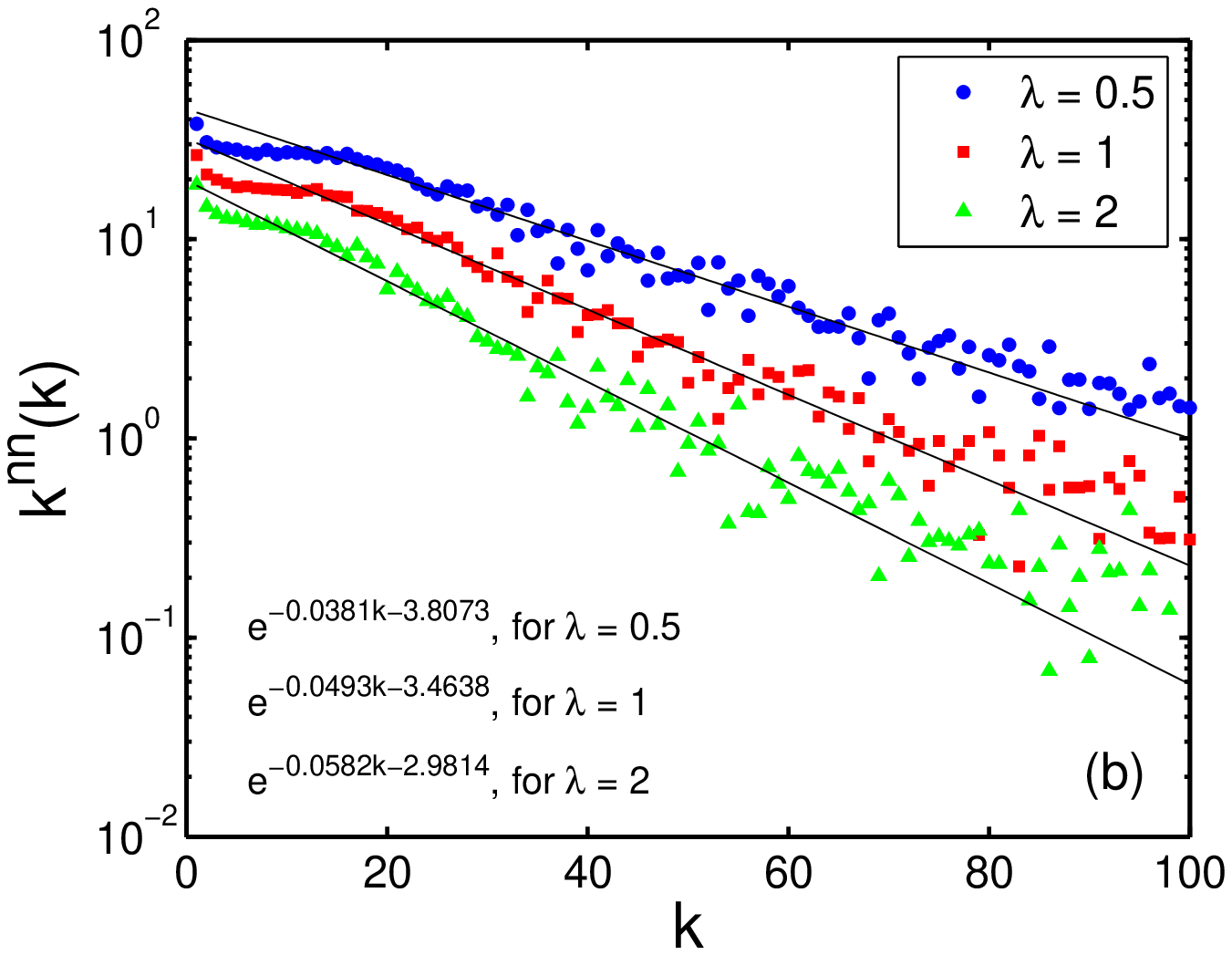} \\
 \end{tabular}
 \end{center}
 \caption{(a) Clustering coefficient $c(k)$ and (b) average nearest
 neighbor degree $k^{nn}$, for $\lambda = 0.5 ,1 ,2$ and a fixed
 $\alpha_A=\alpha_M=0$, for $N=1000$. Averaging was done over 100 realizations.
 } 
 \label{cc-knn-lamda}
\end{figure}

We now turn to the clustering and neighbor connectivity of the
emerging networks. In Fig. \ref{cc-knn-lamda} we show the
clustering coefficient $c$ and the average neighbor connectivity $k^{nn}$
as a function of $k$. For both quantities, the functional form of the decline  with $k$
is well fit with a 2-parameter exponential fit,
$\exp(-\epsilon_1 \, k + \epsilon_2)$. 

\begin{figure}[t]
 \begin{center}
 \begin{tabular}{cc}
 {\Large clustering} & {\Large neighbor conn.}\\
 \includegraphics[height=60mm]{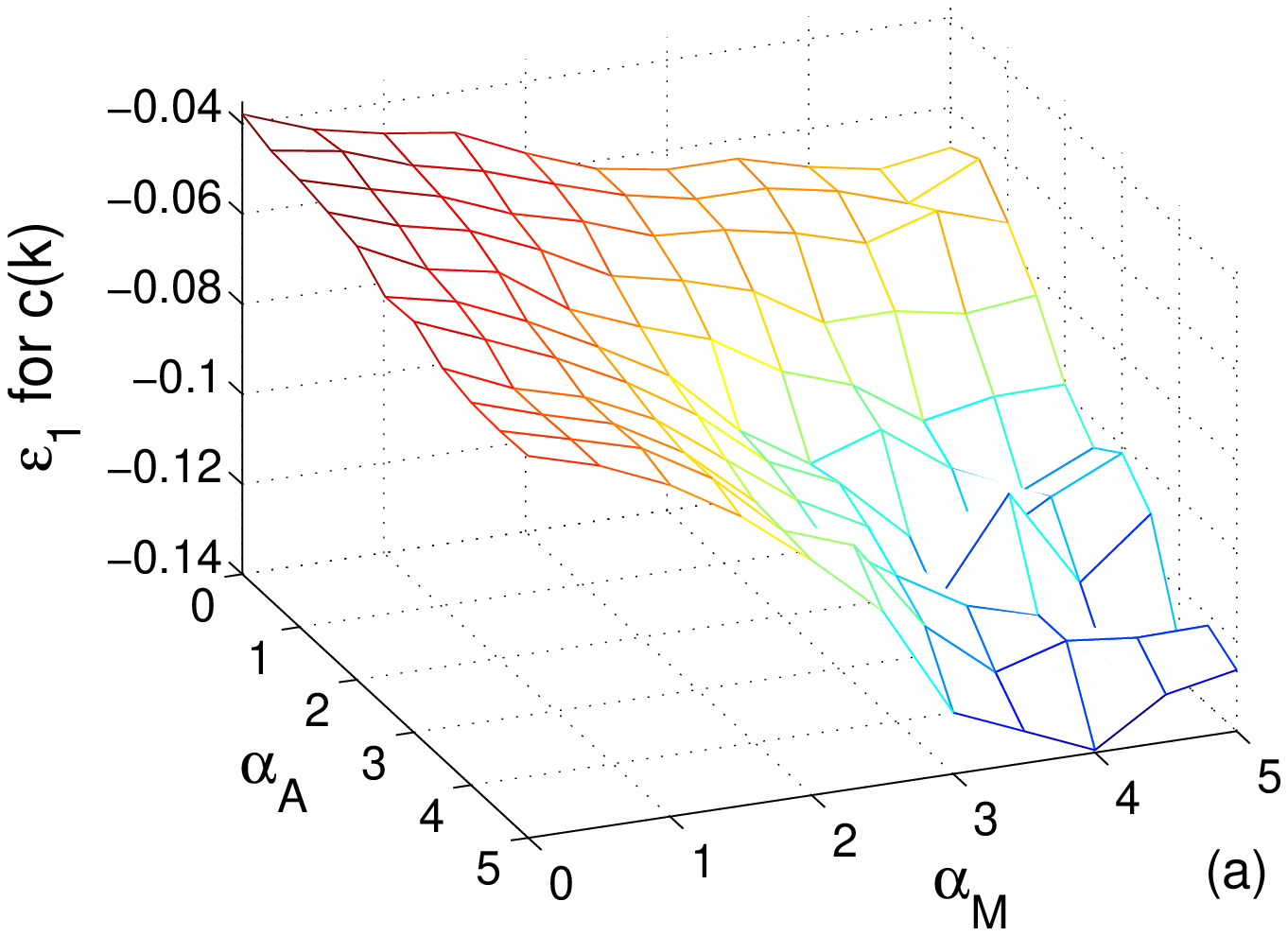}&
 \includegraphics[height=60mm]{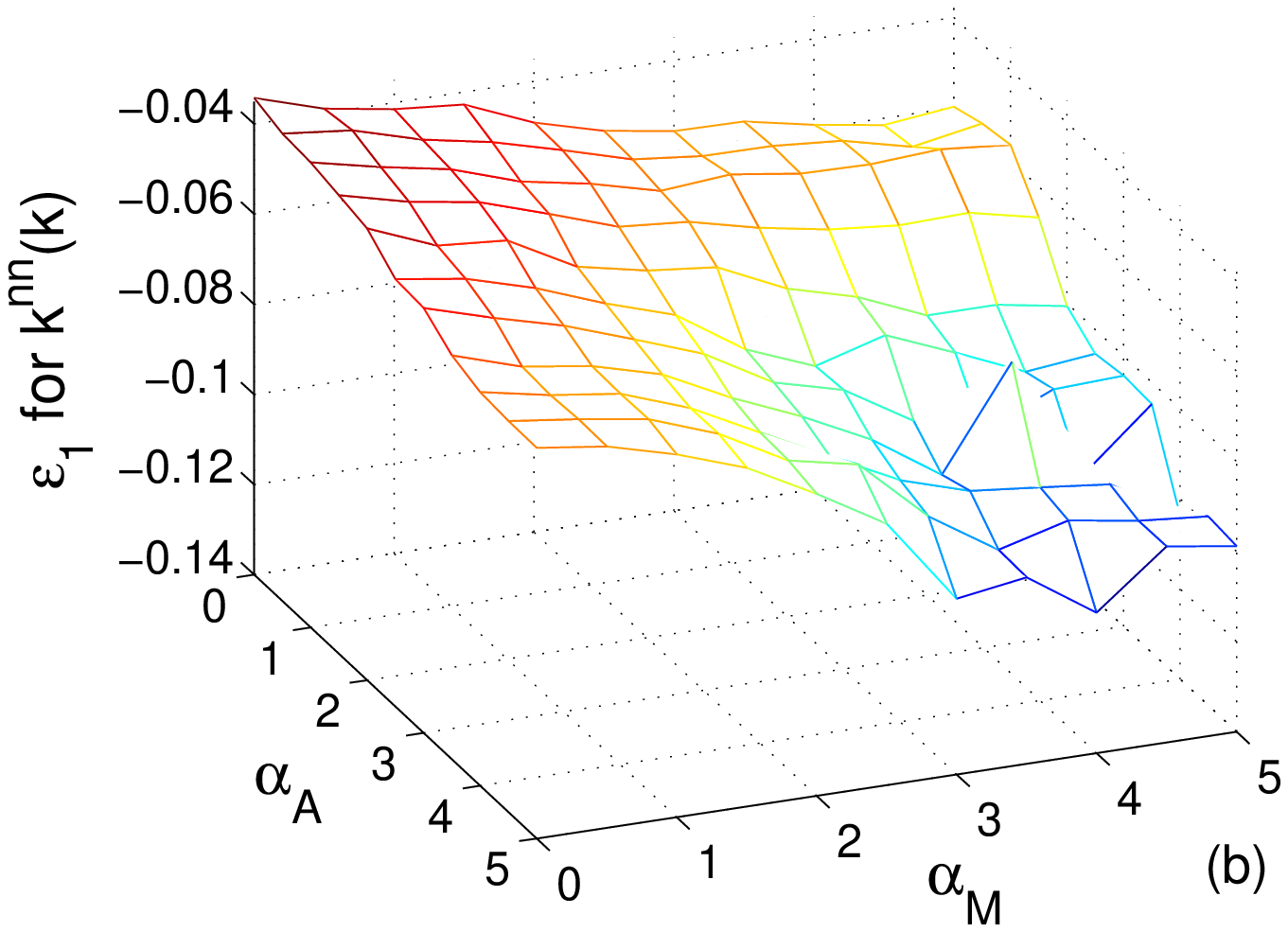}\\
 \end{tabular}
 \end{center}
 \caption{Exponential decay constants $\epsilon_1$  for
 $c(k)$ (a) and $k^{nn}(k)$ (b) over $\alpha_A$ and $\alpha_M$
 for $\lambda = 0.5$. The fit range was  $k \in [1,100]$ and averages over 
 100 independent configurations have been taken. Fits for $\alpha_A>5$ and $\alpha_M>5$
 become statistically insignificant.}
 \label{3d-cc-knn}
\end{figure}
In Fig. \ref{3d-cc-knn} we show the fit parameters $\epsilon_1$
for $c(k)$, (a), and $k^{nn}(k)$, (b), for $\lambda=0.5$. 
For larger $\lambda$ the clustering coefficients become 
drastically smaller, as expected for the $\lambda \to \infty$ and $\alpha_A \to 0$ limit.
Fits for $\alpha_A>5$ and $\alpha_M>5$
become increasingly noisy and are omitted from the figure. 

\begin{figure}[t]
\begin{center}
\begin{tabular}{cc}
{\Large Model} & {\Large Random} \\
\includegraphics[height=50mm]{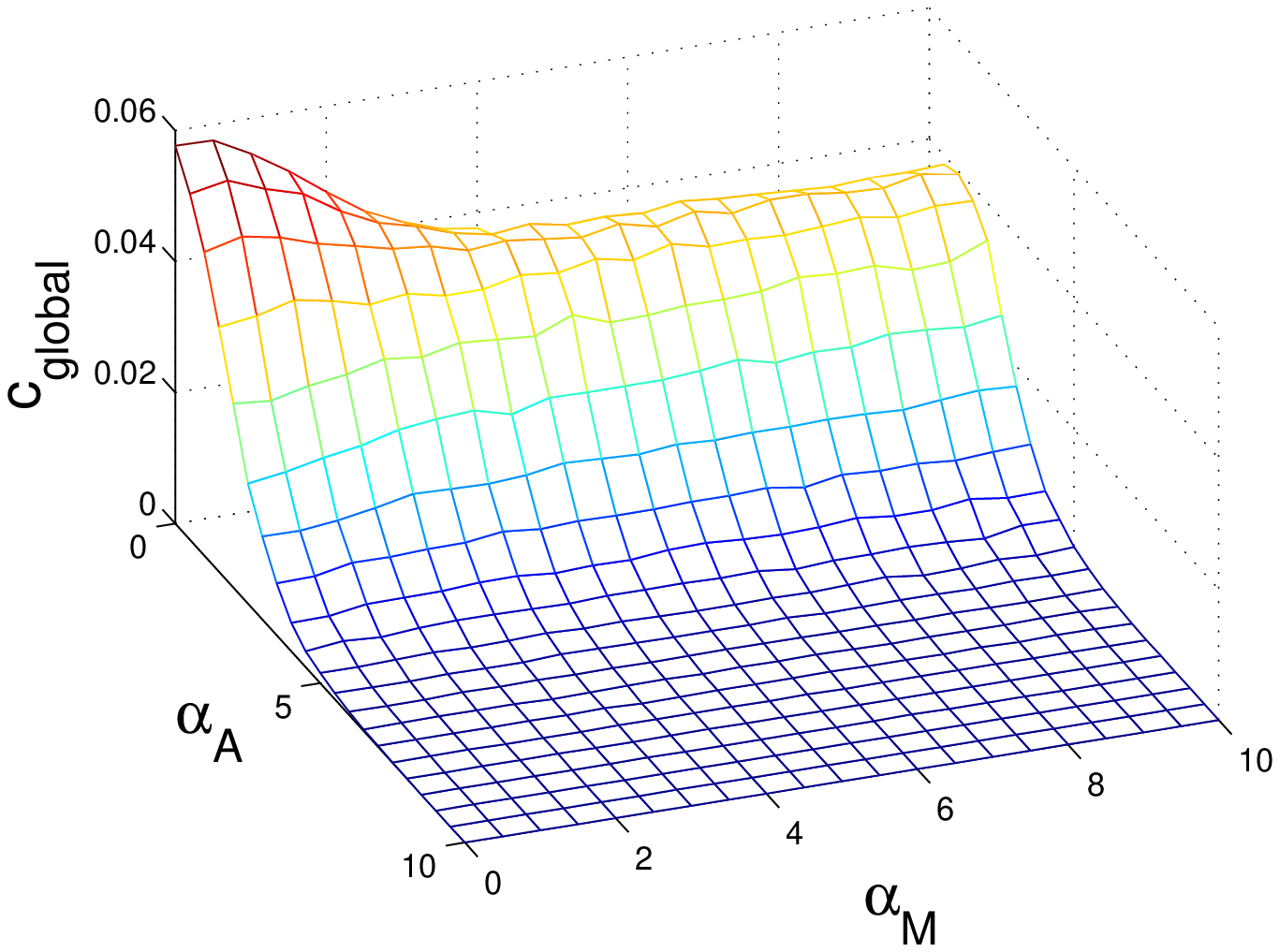}&
\includegraphics[height=50mm]{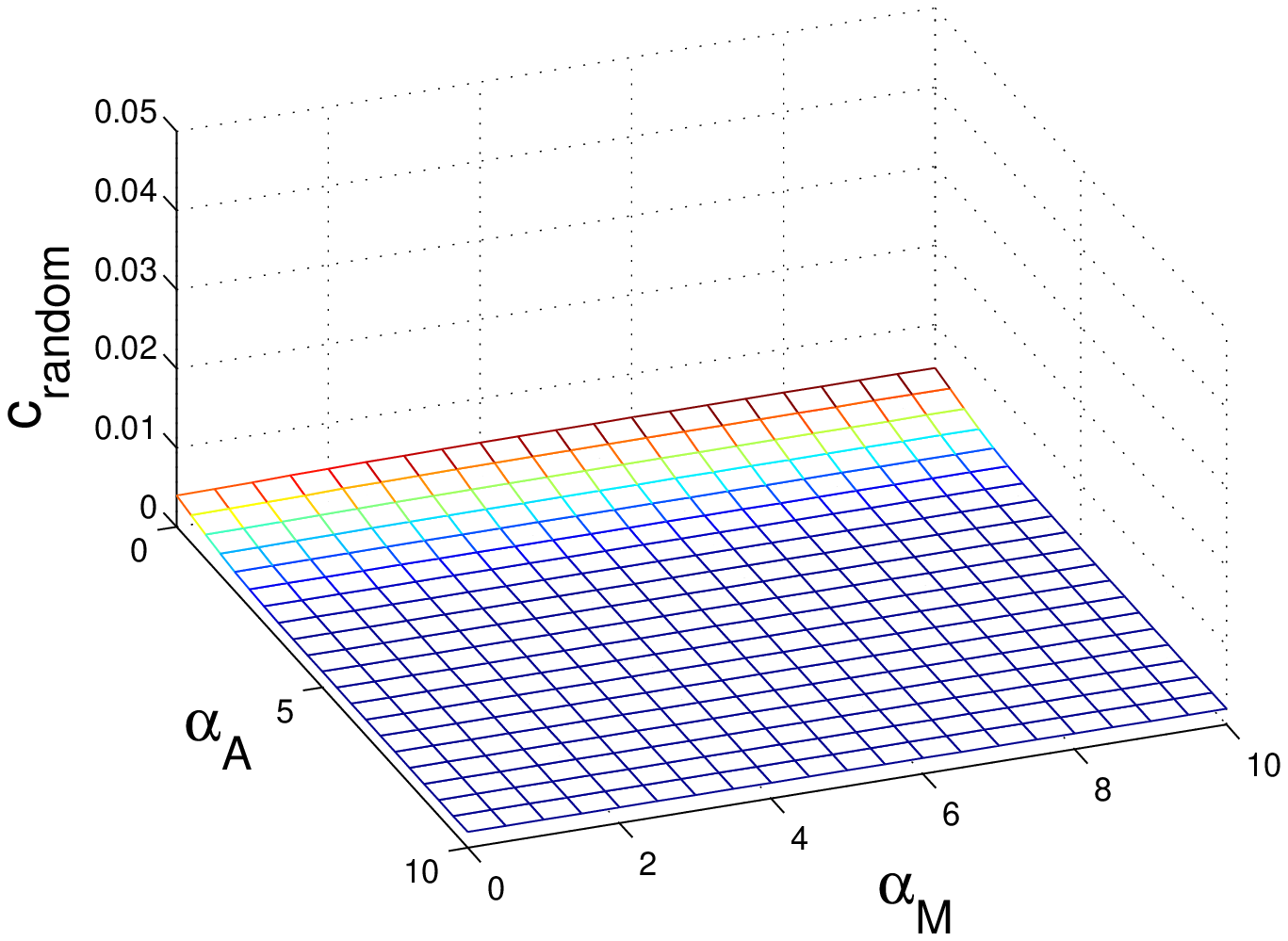}\\
\includegraphics[height=50mm]{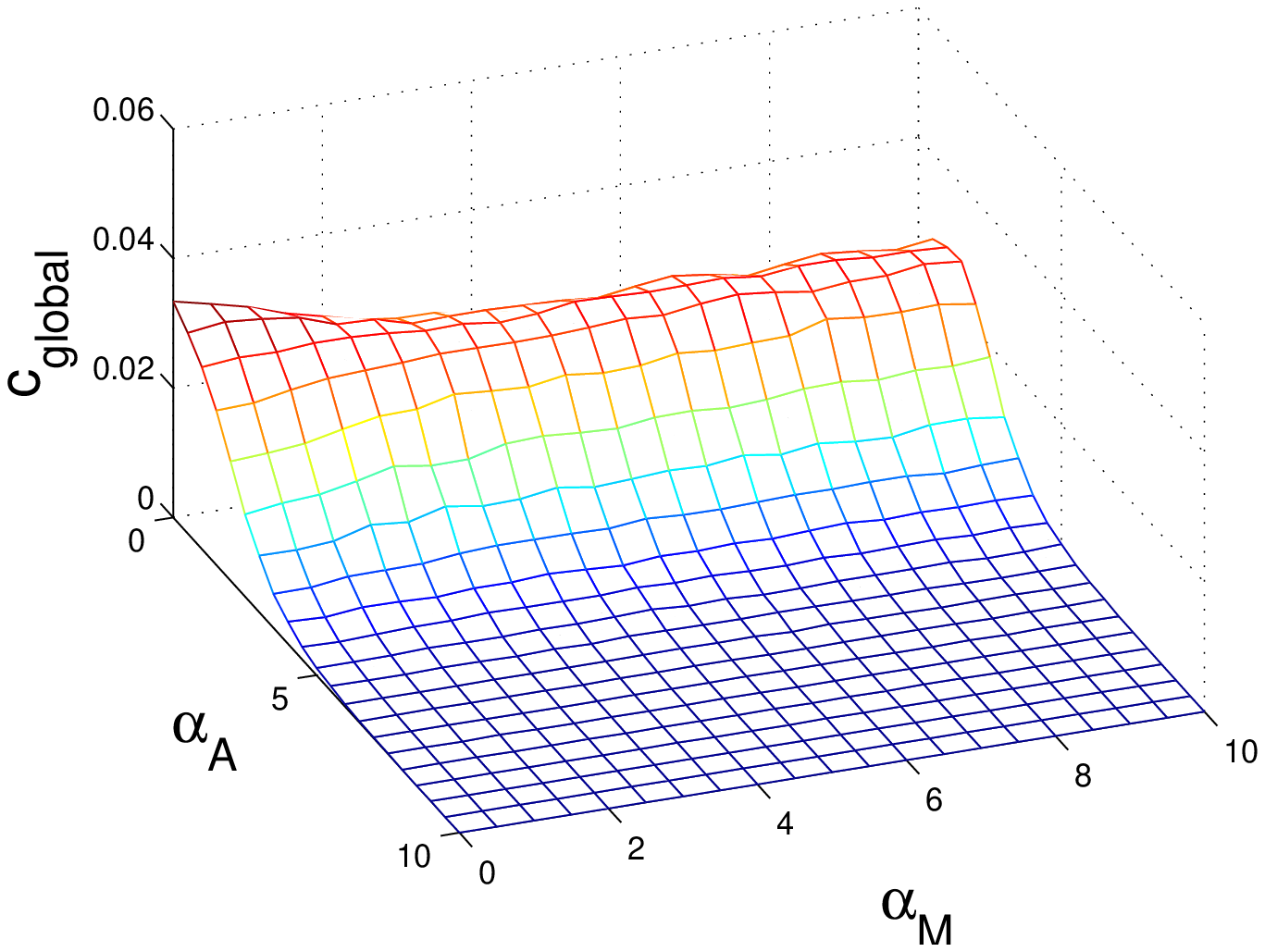}&
\includegraphics[height=50mm]{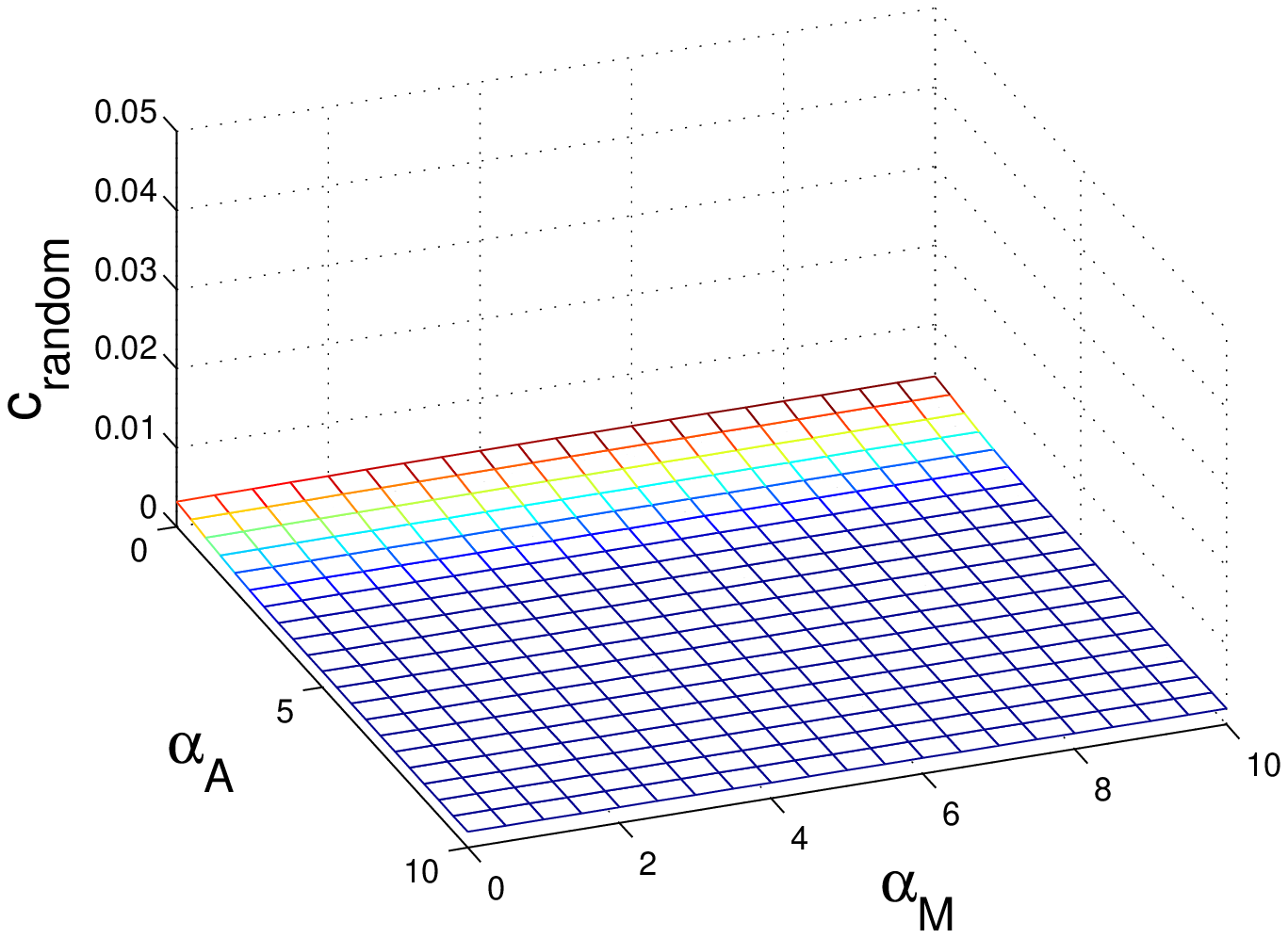}\\
\includegraphics[height=50mm]{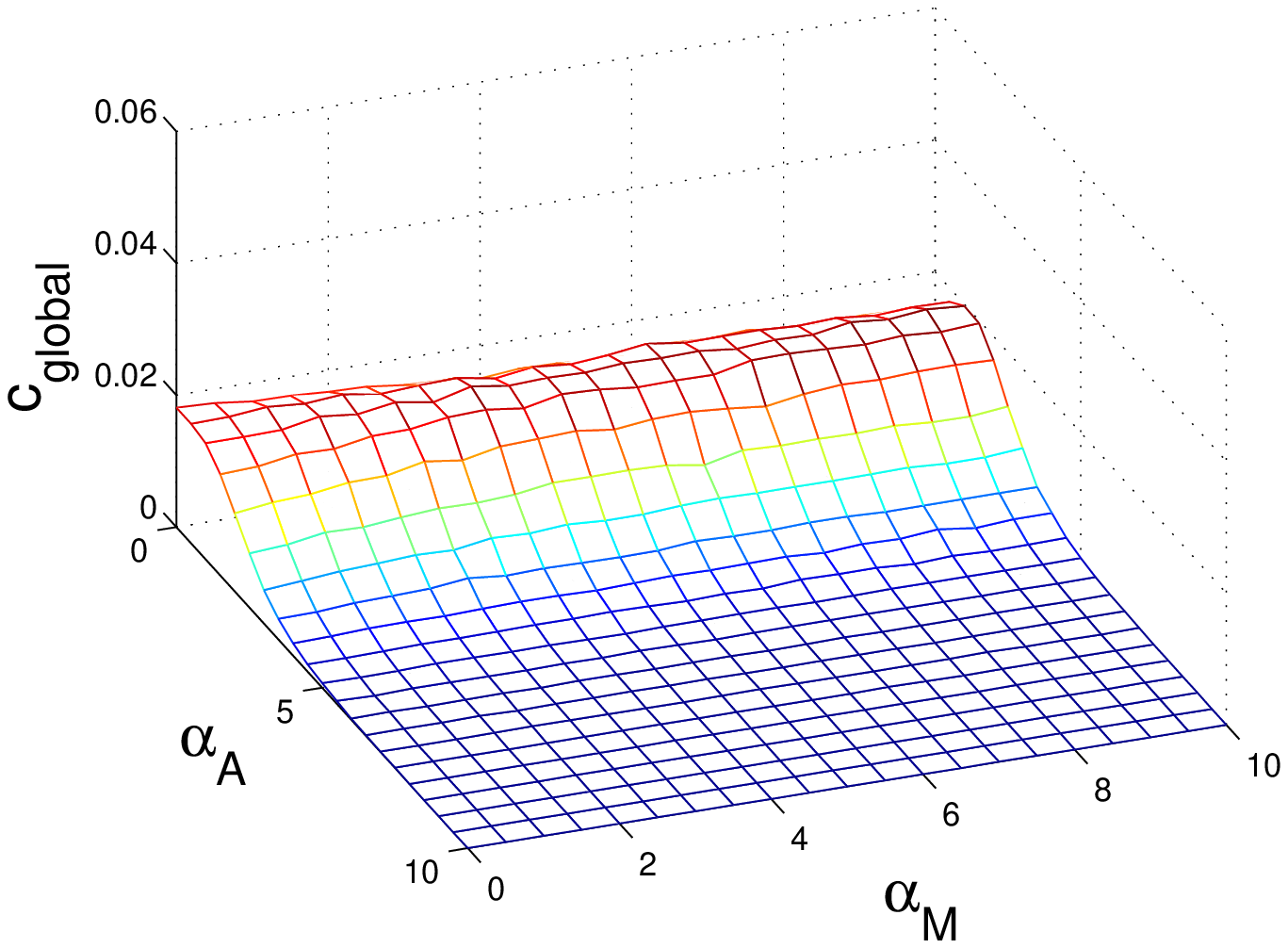}&
\includegraphics[height=50mm]{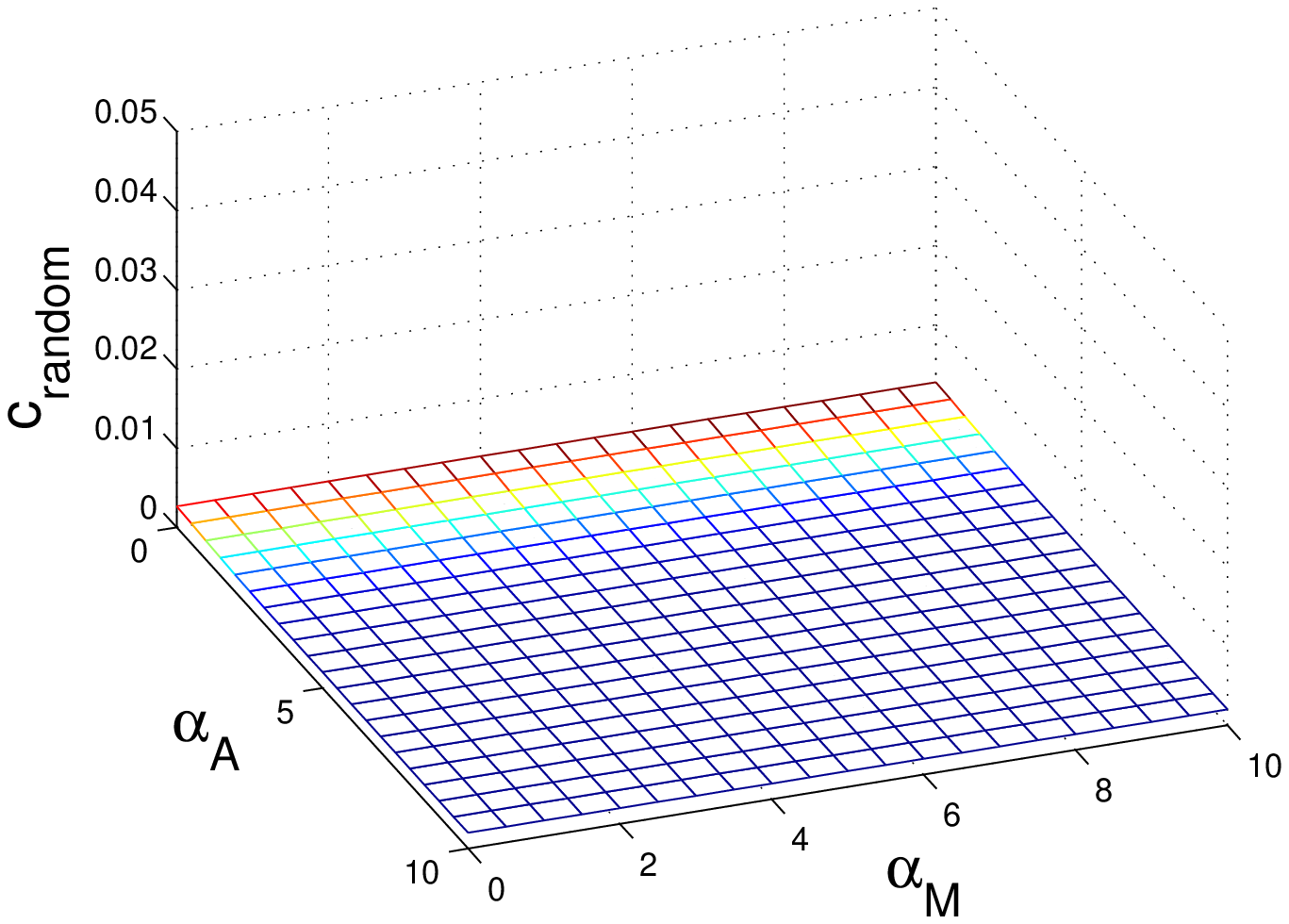}\\
\end{tabular}
\end{center}
\caption{Global clustering coefficient  in the  $\alpha_A$--$\alpha_M$ plane 
 for  $\lambda=0.5,1,2$ (from top to bottom) and $N=1000$ for the present model (a) and for an ER random graph with the same number of links and nodes (b). Averages were taken over  100 independent realizations. 
}
 \label{cglob}
\end{figure}

In Fig. \ref{cglob} we compare the global clustering 
coefficients from our model, with those obtained from a random graph with 
the same dimensions (same number of nodes and links). For the 
Erd\"os-Renyi random graph the clustering coefficient is 
$C_{\rm rand}=\langle k \rangle / N-1$.
The comparison makes  clear that there is almost 
no attachment effects for $\alpha_A>3$ (i.e., negligible dependence from $\alpha_A$), 
and a strong dependence on $\alpha_M$ and $\lambda$, as expected.

\section{Discussion}

We have introduced a general network formation model which is able
to recover, as particular instances, a large class of known network types. We checked that, 
to a very good approximation, the resulting degree distributions
exhibit  $q$-exponential forms, with $q > 1$. While a full theory of
how complex networks are connected to $q \neq 1$ statistical
mechanics is still missing, we provide further evidence that such a
relation does indeed exist. For example, if we associate a finite fixed energy or ``cost" to every bond, and associate with each node half of the energy corresponding to its bonds (the other half corresponding to the other nodes linked by those same bonds), then the degree distribution can be seen as an energy distribution of the type emerging within nonextensive statistical mechanics. It might well be that the full understanding of
this relation  arises from the discrete nature of networks. The
importance of appropriate values of $q \neq 1$ for systems 'living'
in topologies with a vanishing Lebesgue measure  has been pointed out
before \cite{gellmann}. This  possibly makes phase space for certain
nonextensive systems look like a network itself. In this view the
basis of nonextensive systems could be related to a network-like
structure of their 'phase space', explaining the ubiquity of
$q$-exponential distribution functions in the world of networks.

Let us end by pointing out that, in variance with frequent such statements in the 
literature, the present model neatly illustrates that never ending growth is {\it not} 
necessary for having networks that are (asymptotically) scale-free. Indeed, 
$q$-exponential degree distributions do emerge for large enough networks 
which do {\it not} necessarily keep growing.

\begin{acknowledgments}
S.T. acknowledges support from the Austrian Science Foundation 
FWF project  P19132 and would like to thank the SFI and in particular J.D. Farmer for
their great hospitality and support in  2005, where this work was
initiated. C.T. acknowledges partial support from Pronex, CNPq and Faperj (Brazilian agencies).
\end{acknowledgments}

\end{document}